\documentclass[%
 reprint,
 amsmath,amssymb,
 aps,
]{revtex4-2} 

\usepackage{graphicx}
\usepackage{listings}

\usepackage{dcolumn}
\usepackage{bm}
\usepackage[hidelinks]{hyperref}
\usepackage{braket}
\usepackage{comment}
\usepackage[caption=false]{subfig}
\usepackage{adjustbox}
\usepackage{placeins}

\begin{document}

\preprint{APS/123-QED}

\title{A Hybrid Transformer Architecture with a Quantized Self-Attention Mechanism Applied to Molecular Generation}
\thanks{Contact author: \\anthony.smaldone@yale.edu\\ victor.batista@yale.edu}%

\author{Anthony M. Smaldone$^1$, Yu Shee$^1$, Gregory W. Kyro$^1$, Marwa H. Farag$^2$, Zohim Chandani$^3$, Elica Kyoseva$^2$, and Victor S. Batista$^{1,4}$}

\affiliation{$^1$Department of Chemistry, Yale University, New Haven, CT 06520, USA \\
$^2$NVIDIA Corporation, 2788 San Tomas Expressway, Santa Clara, 95051, CA, USA \\
$^3$NVIDIA Corporation, Quantum Algorithm Engineering, London, UK\\
$^4$Yale Quantum Institute, New Haven, CT 06511, USA}






\date{\today}

\begin{abstract}
The success of the self-attention mechanism in classical machine learning models has inspired the development of quantum analogs aimed at reducing computational overhead. Self-attention integrates learnable \textit{query} and \textit{key} matrices to calculate attention scores between all pairs of tokens in a sequence. These scores are then multiplied by a learnable \textit{value} matrix to obtain the output self-attention matrix, enabling the model to effectively capture long-range dependencies within the input sequence. Here, we propose a hybrid quantum-classical self-attention mechanism as part of a transformer decoder, the architecture underlying large language models (LLMs). To demonstrate its utility in chemistry, we train this model on the QM9 dataset for conditional generation, using SMILES strings as input, each labeled with a set of physicochemical properties that serve as conditions during inference. Our theoretical analysis shows that the time complexity of the \textit{query-key} dot product is reduced from $\mathcal{O}(n^2 d)$ in a classical model to $\mathcal{O}(n^2\log d)$ in our quantum model, where $n$  and $d$ represent the sequence length and embedding dimension, respectively. We perform simulations using NVIDIA's CUDA-Q platform, which is designed for efficient GPU scalability. This work provides a promising avenue for quantum-enhanced natural language processing (NLP).

\end{abstract}

\maketitle


\begin{figure*}[!htbp]
    \centering
    \includegraphics[width=1.0\textwidth]{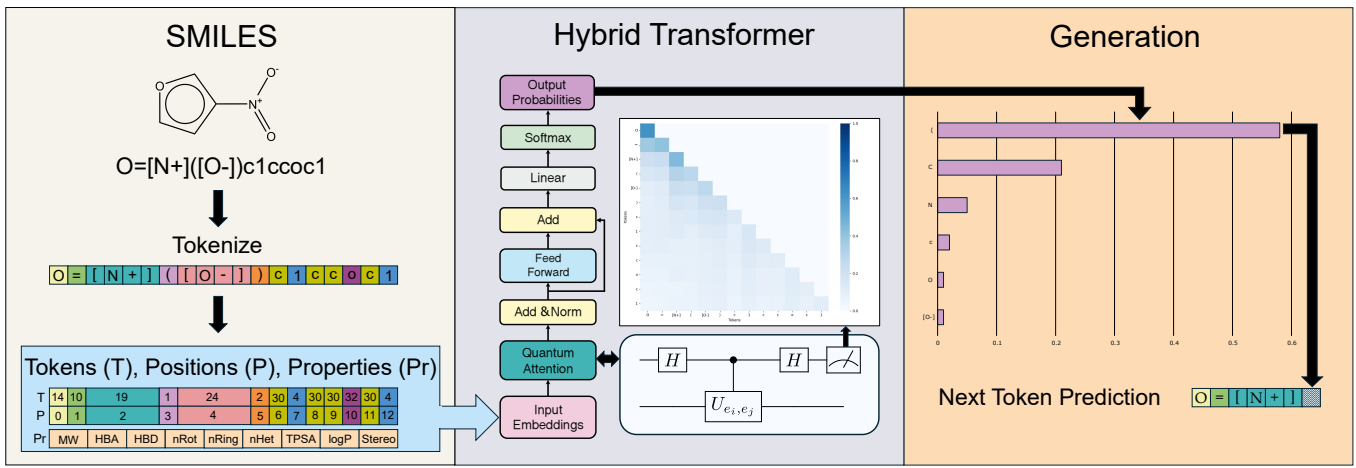}
    \caption{The proposed hybrid quantum-classical transformer model generates molecules by processing SMILES strings (e.g., O=[N+]([O-])c1ccccc1). Each string is split into a sequence of tokens, which are assigned token, positional, and physicochemical property embeddings. These embeddings pass through a hybrid self-attention mechanism: quantum circuits compute attention scores, which are combined with classical value matrices. The output then flows through the remaining classical transformer decoder to predict the next token in the sequence. This enables conditional molecular generation targeting specific physicochemical properties.}
    \label{fig:abstract_figure}
\end{figure*}

\section{\label{sec:level1}Introduction}
\subsection{Motivation}
The self-attention mechanism, a cornerstone of the Transformer architecture \cite{vaswani_attention_2017}, has revolutionized numerous machine learning (ML) domains, including natural language processing (NLP) \cite{openai_gpt-4_2024}, computer vision \cite{dosovitskiy_image_2021, carion_end--end_2020}, and computational biology \cite{jumper_highly_2021, abramson_accurate_2024}. By capturing long-range dependencies in sequential data, it enables efficient and scalable learning, driving the Transformer’s widespread adoption. Its versatility has spurred extensive research into refining and extending its applications within and beyond this framework.

Quantum machine learning (QML) has emerged as a rapidly growing field \cite{biamonte_quantum_2017,wiebe_quantum_2012,lloyd_quantum_2014,rebentrost_quantum_2014, cong_quantum_2019, zoufal_quantum_2019}, leveraging quantum computation to potentially enhance learning and optimization tasks. This field explores whether manipulating quantum states in Hilbert space outperforms classical vector operations in deep learning. Inspired by the success of the self-attention mechanism and the Transformer architecture, researchers are increasingly exploring quantum analogs to investigate potential performance gains achievable through the learning of information encoded into quantum states.  Recently, Loshchilov et al. \cite{loshchilov_ngpt_2024} introduced a normalized transformer with representation learning on a hypersphere. This approach bears similarity to quantum state evolution, where unitary operators move states across a hypersphere, suggesting that high-dimensional normalized representations may offer advantages for quantum self-attention mechanisms.

\subsection{Background}

The earliest application of self-attention in QML came from Li et al. \cite{li_quantum_2023}, who used classical Gaussian projections of query and key quantum states for text classification. Some works depart from the classical formulation of the scaled dot-product attention mechanism and ``mix" tokens together in Hilbert space to capture correlations instead of computing query-key dot products. For instance, Khatri et al. \cite{khatri_quixer_2024} develop a quantum algorithm of the skip-k-gram NLP technique using linear combinations of unitaries (LCU) and the quantum singular value transform (QSVT). Zheng et al. \cite{zheng_design_2023} encode both query and key vectors into a parametric quantum circuit (PQC) and measure the qubits to learn their correlations. Evans et al. \cite{evans_learning_2024} replace the explicit dot product with a PQC that blends tokens in the Fourier domain via quantum Fourier transformers (QFTs).

Other efforts focus on quantum analogs of self-attention and transformers that closely adhere to the classical framework, preserving the core principles of their operation. Xue et al. \cite{xue_end--end_2024} propose an end-to-end quantum vision transformer; however, reliance of analog encoding for quantum random access memory (qRAM), leads to exponential scaling unless binary-tree structured data is assumed \cite{mitarai_quantum_2019}, limiting its feasibility for noisy intermediate-scale quantum (NISQ) and general-purpose applications. Cherrat et al. \cite{cherrat_quantum_2024} introduce a hybrid approach, learning query and key states with $\mathcal{O}(d)$ qubits--where $d$ represents the embedding dimension--to compute the squared dot product as an attention score. Meanwhile, Liao and Ferrie \cite{liao_gpt_2024} and Guo et al. \cite{guo_quantum_2024} propose theoretical quantum Transformer models based on algorithms like LCU and block encoding to utilize quantum linear algebra techniques. While these methods offer improved scaling under sparsity assumptions, their quantum resource requirement continues to make them impractical for the NISQ era, underscoring the demand for NISQ-friendly quantum self-attention approaches.

In this work, we introduce a novel quantum-classical hybrid self-attention mechanism, integrated into a transformer decoder for molecular generation. This approach uses $\mathcal{O}(\log{d})$ qubits and CNOT gates to learn all embeddings, as well as query and key representations, quantum mechanically. Unlike prior methods, it directly yields attention scores without squaring the dot product. We further incorporate positional embeddings and establish a general framework for additional embeddings, such as physicochemical molecular features, enabling control over generated molecular properties. Our results demonstrate that this hybrid model performs on par with classical baselines in SMILES validity, uniqueness, novelty, and property-targeted molecular generation.

\section{Framework}
\subsection{Classical Attention Score Calculation}
For a given input sequence of tokens $\{x_1, x_2, \dots, x_n\}$, each token $x_i$ is mapped to an embedding $\mathbf{e}_i$ via a learned embedding matrix. Positional embeddings $\mathbf{p}_i$ are added to preserve token order, yielding the final input embeddings:
\begin{eqnarray}
    \mathbf{z}_i = \mathbf{e}_i + \mathbf{p}_i,
\end{eqnarray}
where the embedding dimension is $d$ (i.e., $\dim \mathbf{e}_i = \dim \mathbf{p}_i = d$). Additional embeddings can enhance next-token prediction and condition the model to generate data with specific properties during inference. These are incorporated by summing $\kappa$ additional vectors $\mathbf{c}_{i,v}$ with the token and positional embeddings, as in:
\begin{eqnarray}
    \mathbf{z}_i = \mathbf{e}_i + \mathbf{p}_i + \sum^\kappa_{v=1}\mathbf{c}_{i_v}
    \label{equation:classically_including_conditions}
\end{eqnarray}
following established practice \cite{devlin_bert_2019}.

The input embeddings are then linearly projected into \textit{query} $(\mathbf{Q})$, \textit{key} $(\mathbf{K})$, and \textit{value} $(\mathbf{V})$ matrices:

\begin{eqnarray}
    \mathbf{Q} = \mathbf{Z} \mathbf{W}^Q, \quad \mathbf{K} = \mathbf{Z} \mathbf{W}^K, \quad \mathbf{V} = \mathbf{Z} \mathbf{W}^V,
    \label{eq:classical_Q_K_V}
\end{eqnarray}
where $\mathbf{Z} = [\mathbf{z}_1, \mathbf{z}_2, \dots, \mathbf{z}_n]^\top$ stacks the input embeddings, and $\mathbf{W}^Q$, $\mathbf{W}^K$, and $\mathbf{W}^V$ are learned weight matrices. The scaled dot-product attention mechanism \cite{vaswani_attention_2017} computes attention scores across all token pairs as:

\begin{equation}
    \mathbf{Attention(Q,K,V)}= \mathbf{A}(\mathbf{Q},\mathbf{K}) \mathbf{V},
\end{equation}
with
\begin{eqnarray}
    \mathbf{A (Q,K)} = \text{softmax}\left( \frac{\mathbf{Q} \mathbf{K}^\top}{\sqrt{d_k}} \right)
    \label{equation:classical_attention_matrix}
\end{eqnarray}
where $d_k$ is the dimension of the key vectors.

\subsection{\label{sec:learning_attention_scores}Learning Attention Scores with Quantum States}
\begin{figure}[!htbp]
    \centering
    \includegraphics[width=0.35\textwidth]{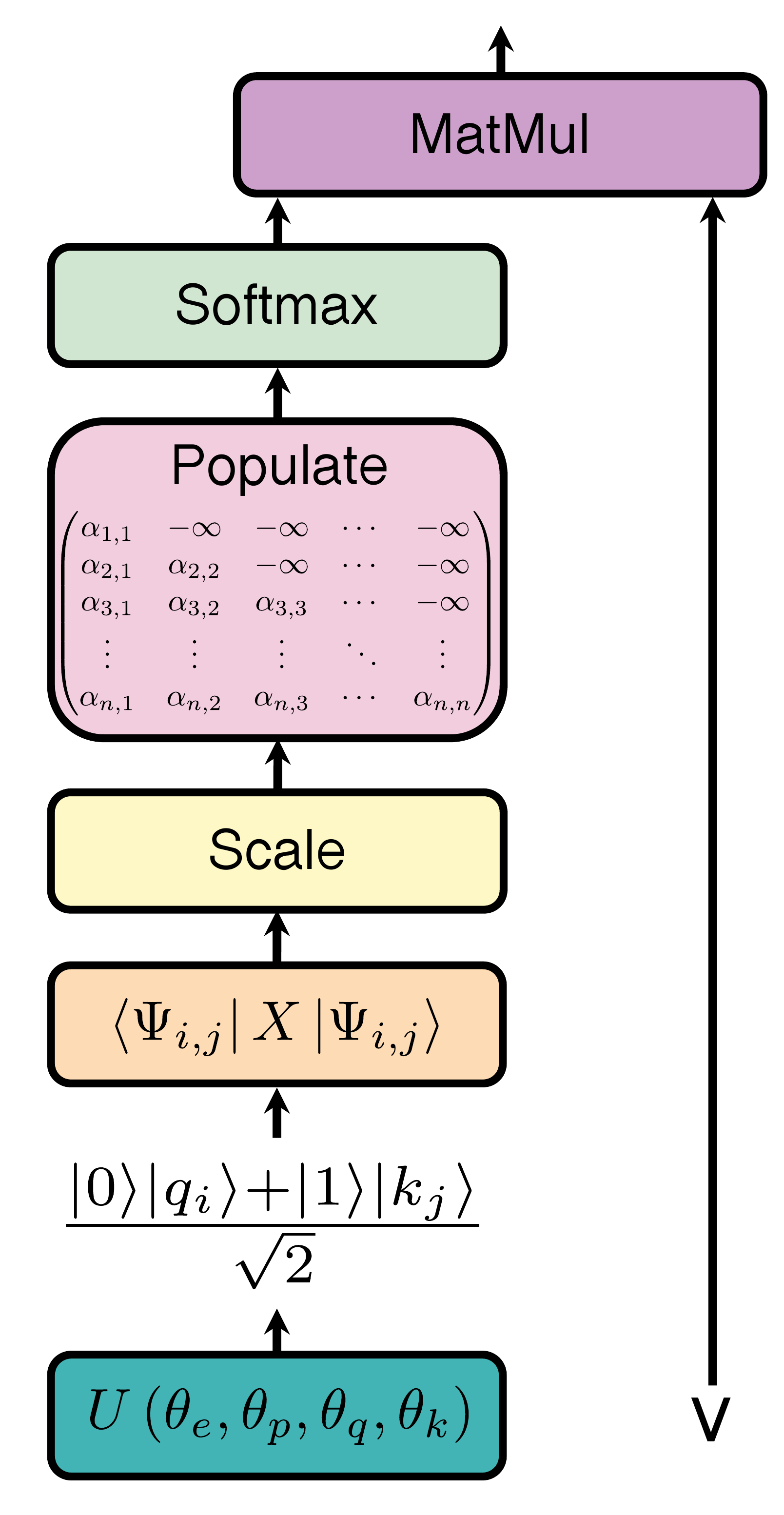}
    \caption{Quantum self-attention layer combining $\mathbf{QK}^\top$ calculated with quantum circuits and a classically computed value matrix $\mathbf{V}$ (Equation \ref{eq:classical_Q_K_V}). Quantum token embeddings $\mathbf{\theta_e}$, positional angles $\mathbf{\theta_p}$, and learnable parameters $\mathbf{\theta_q}$, $\mathbf{\theta_k}$ are used in the unitary $U$ as the circuit that evolves the quantum states depicted in Equations \ref{eq:psi_1} -- \ref{eq:psi_6}. Each quantum circuit produces an attention score, thus there are $\frac{n^2+n}{2}$ instances of $U$ with their respective angles to ensure a fully populated masked attention matrix. The expectation value of the Pauli-X observable on the ancilla qubit (equivalent to a Hadamard transform and measurement in the computational basis as shown in Equations \ref{eq:final_state} and \ref{eq:measurement}) is obtained and represents the dot product between query and key vectors. The original transformer implementation \cite{vaswani_attention_2017} scales attention scores by $\frac{1}{\sqrt{d_k}}$ to maintain a variance of 1. Since the dot products herein are obtained from the expectations of the quantum subsystems, they are bound on the closed interval $[-1,1]$. To maintain a variance of 1, they must be scaled by $\sqrt{d_k}$. The scaled values are stored in the masked attention matrix, softmax is applied to the rows, and the resulting matrix is multiplied with $\mathbf{V}$.}
    \label{fig:quantum_self_attention_architecture}
\end{figure}

In the attention matrix $\mathbf{A}$, each element $a_{i,j}$ is the scaled dot product of the $i$-th query vector $\mathbf{q}_i$ and the $j$-th key vector $\mathbf{k}_j$, followed by a softmax operation (see Equation \ref{equation:classical_attention_matrix}). In this work, we use quantum circuits to compute individual attention scores. We learn representations of the embedding vectors $\mathbf{z}_i$, query vectors $\mathbf{q}_i$, and key vectors $\mathbf{k}_i$ as quantum states, denoted $\ket{q_i}$ and $\ket{k_j}$, and determine their inner product $\braket{q_i | k_j}$. The value matrix $\mathbf{V}$ and subsequent operations, however, remain classical. Figure \ref{fig:quantum_self_attention_architecture} illustrates our hybrid quantum-classical self-attention framework. The next subsection details the quantum circuit used to compute $\mathbf{A}$’s matrix elements.

\subsubsection{Quantum Encoding of Token and Positional Information}

Similar to the classical framework, we construct an embedding matrix to assign token embeddings $\mathbf{\theta_{e_i}}$, where each vector’s entries are scaled to $[0, \pi]$ and its dimension equals the number of learnable parameters $m$ in the ansatz $U_e$, which prepares the quantum state $\ket{e_i}$ for each token. In this work, we define learnable positional encoding angles $\mathbf{\theta_{p_i}}$, initialized to zero. The states $\ket{e_i}$ and $\ket{p_i}$ are prepared by applying unitaries $U_{e_i}=U_{e}\left(\mathbf{\theta_{e_i}}\right)$ and $U_{p_i}=U_{p}\left(\mathbf{\theta_{p_i}}\right)$ to initial states:

\begin{eqnarray}
    U_{e}\left(\mathbf{\theta_{e_i}}\right)\ket{0}^{\otimes \frac{\log{d}}{2}} = \ket{e_i}, U_{p}\left(\mathbf{\theta_{p_i}}\right)\ket{0}^{\otimes \frac{\log{d}}{2}} = \ket{p_i}.
    \label{equation:token_and_pos_unitaries}
\end{eqnarray}

\begin{figure*}[!htbp]
    \centering
    \includegraphics[width=1.0\textwidth]{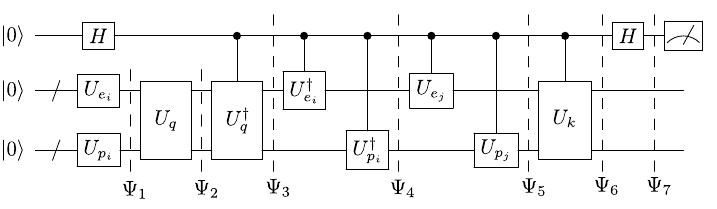}
    \caption{Quantum circuit used to create query and key states from a given quantum token and positional encoding to produce an attention score when measured. $U_{e_i}/U_{e_j}$ and $U_{p_i}/U_{p_j}$ are the unitaries that create the token and positional encoding of the $i$/$j$-th token into the quantum state. $U_{q}$ and $U_{k}$ are the unitaries to learn query and key representations of the quantum states containing token and positional information. The mathematical description of states $\Psi_1$ to $\Psi_7$ are found in Equations \ref{eq:psi_1} to \ref{eq:final_state}. The expectation value on the ancilla qubit yields the desired query-key dot product $\text{Re}\braket{q_i|k_j}$.}
    \label{fig:circuit_sequence_only}
\end{figure*}

 Just as the elements of the token and positional embeddings are learned classically, the parameters of the unitary operators $U_{e_i}$ and $U_{p_i}$ are learned as well. These states are prepared independently, yielding the composite state $\Psi_1$ as shown in Figure \ref{fig:circuit_sequence_only} which encodes token and positional information for a sequence:

\begin{eqnarray}
 \Psi_1 = \ket{e_i} \otimes \ket{p_i} = \ket{z_i}.
 \label{eq:psi_1}
\end{eqnarray}

\begin{figure}[!htbp]
    \centering
    \includegraphics[width=0.25\textwidth]{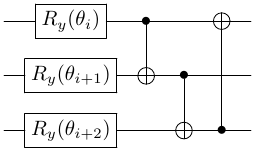}
    \caption{Structure of the parametric quantum circuits used in this work.}
    \label{fig:pqc}
\end{figure}

All PQC ansatzes in this work use a single layer of $R_y$ gates followed by an entangling layer of CNOT gates as shown in Figure \ref{fig:pqc}.

\subsubsection{Learning Query and Key States}

The separable quantum states of token and positional encodings are entangled via a PQC $U_{q}$, resulting in a quantum state analogous to a query vector. 
\begin{eqnarray}
\Psi_2 = U_{q} \ket{z_i} = \ket{q_i}
\end{eqnarray}

An ancilla qubit in the $\ket{+}$ state is introduced, and the entire circuit is conditionally reversed under its control. If the ancilla qubit is in the $\ket{0}$ state, the working register remains $\ket{q_i}$. If the ancilla qubit is in the $\ket{1}$ state, the working register is reset to $\ket{0}^{\otimes \log{d}}$ as shown in Equations \ref{eq:controlled_adjoint_circuit_1} and \ref{eq:controlled_adjoint_circuit_2}

\begin{eqnarray}
    \Psi_3 = CU^{\dagger}_{q}\left(\frac{\ket{0}+\ket{1}}{\sqrt{2}} \otimes \ket{q_i}\right) = \frac{\ket{0}\ket{q_i} + \ket{1}\ket{z_i}}{\sqrt{2}}
    \label{eq:controlled_adjoint_circuit_1}
\end{eqnarray}

\begin{eqnarray}
    \Psi_4 = CU^{\dagger}_{e_i}\left(CU^{\dagger}_{p_i}\left(\frac{\ket{0}\ket{q_i} + \ket{1}\ket{z_i}}{\sqrt{2}}\right)\right) = \nonumber \\
    \frac{\ket{0}\ket{q_i} + \ket{1}\ket{0}^{\otimes \log{d}}}{\sqrt{2}}
    \label{eq:controlled_adjoint_circuit_2}
\end{eqnarray}

where $CU = \ket{0}\bra{0}\otimes I + \ket{1}\bra{1} \otimes U$ and $I$ is the identity operator.

Analogous to the preparation of $\ket{z_i}$ in Equation \ref{eq:psi_1}, $\ket{z_j}$ is prepared with the difference being all PQCs are controlled by the ancilla qubit. This controlled preparation ensures the modified Hadamard test yields a real-valued dot product $\braket{q_i | k_j}$ between query and key states. After preparing $\ket{z_j}$, a controlled PQC $CU_k$ transforms it into $\ket{k_j}$, applied only when the ancilla is in $\ket{1}$, resulting in the state:

\begin{eqnarray}
\Psi_5= CU_{p_j}\left(CU_{e_j}\left(\frac{\ket{0}\ket{q_i} + \ket{1}\ket{0}^{\otimes \log{d}}}{\sqrt{2}}\right)\right) = \nonumber \\ 
\frac{\ket{0}\ket{q_i} + \ket{1}\ket{z_j}}{\sqrt{2}}
\end{eqnarray}

\begin{eqnarray}
\Psi_6= CU_{k}\left(\frac{\ket{0}\ket{q_i} + \ket{1}\ket{z_j}}{\sqrt{2}}\right) = \nonumber \\ 
\frac{\ket{0}\ket{q_i} + \ket{1}\ket{k_j}}{\sqrt{2}}.
\label{eq:psi_6}
\end{eqnarray}

A final Hadamard gate is applied to the ancilla qubit, which transforms the state to:
\begin{eqnarray}
\Psi_7 = \frac{\ket{0} \otimes \left(\ket{q_i} + \ket{k_j}\right) + \ket{1} \otimes \left(\ket{q_i} - \ket{k_j}\right)}{2}.
\label{eq:final_state}
\end{eqnarray}

The ancilla qubit is measured yielding an expectation value of

\begin{eqnarray}
    \bra{\Psi_7}\left(Z \otimes I^{\otimes \log{d}}\right)\ket{\Psi_7} = \text{Re}\braket{q_i|k_j},
    \label{eq:measurement}
\end{eqnarray}
which is equivalent to the classical analog of the $i,j$-th entry of the $\mathbf{QK}^\top$ matrix, where $Z$ is the Pauli-Z gate.

\begin{figure}[!htbp]
    \centering
    \includegraphics[width=0.40\textwidth]{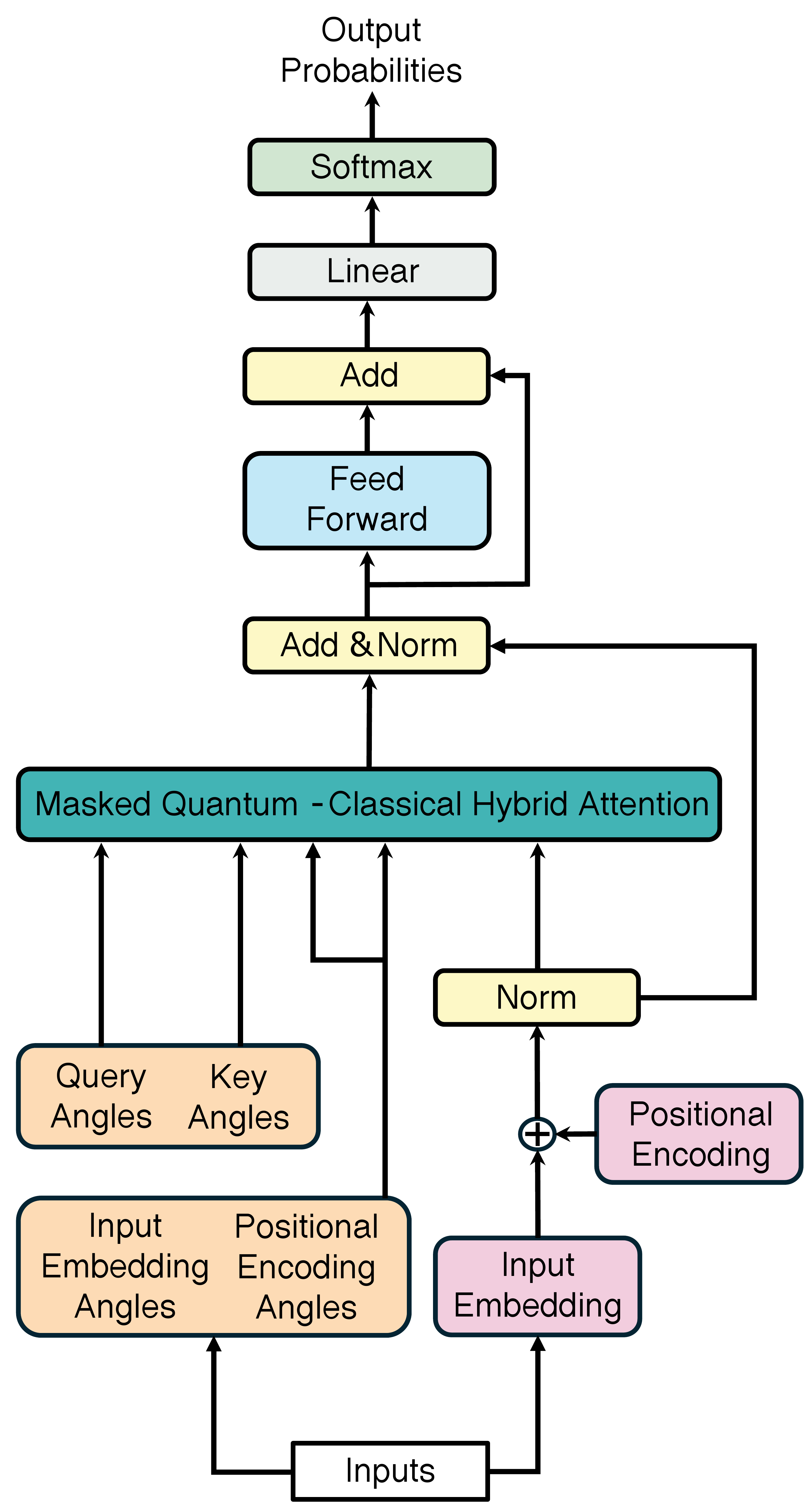}
    \caption{Architecture of the hybrid quantum-classical transformer decoder where embeddings are learned both quantum and classically. $\mathbf{QK}^\top$ is computed with quantum circuits and $\mathbf{V}$ is computed classically.    Embedding and parameter (learnable position) matrices of dimension $n \times \dim{\mathbf{\theta_e}}=n \times \dim{\mathbf{\theta_p}}$ and $n \times d$ are defined for a given input sequence to obtain quantum parameters (orange) and classical (pink) parameters of the model, respectively. The query and key angles are learned and used to transform the quantum embedding states into quantum query and key states. Modified-Hadamard tests are performed, and the output of the masked quantum-classical hybrid attention mechanism (teal) is a matrix of dimensions $n \times d$.}
    \label{fig:overall_quantum_architecture}
\end{figure}

Because each attention score is computed by an independent quantum circuit, this process can be parallelized across multiple quantum processing units (QPUs). For next-token prediction, where the upper triangle of the $\mathbf{Q} \mathbf{K}^\top$ matrix is masked, a maximum of $\frac{n^2 + n}{2} - 1$ unique quantum circuits are required in the worst-case scenario. This ``worst case" occurs because circuits with identical parameters, which produce the same results, need not be recomputed, thus eliminating redundancy. Moreover, the first circuit, corresponding to $a_{1,1}$, need not be executed since its softmax output is always 1 due to masking.

\subsubsection{Extension to Additional Embeddings}

Beyond token and positional encodings, our method can incorporate additional embeddings. To mirror the classical approach of equal embedding dimensions, we choose all Hilbert subspaces for each embedding to be of equal dimension. Thus, to form the quantum state $\ket{\widetilde{z}_i}$, which includes token, positional, and $\kappa$ additional embeddings, we prepare

\begin{eqnarray}
    \ket{\widetilde{z_i}} = U_e\left(\mathbf{\theta_{e_i}}\right)\ket{0}^{\otimes \left(\frac{\log{d}}{\kappa + 2}\right)} \otimes U_p\left(\mathbf{\theta_{p_i}}\right)\ket{0}^{\otimes \left(\frac{\log{d}}{\kappa + 2}\right)} \otimes \nonumber \\
    \bigotimes_{v=1}^\kappa U_{c_v}\left(\mathbf{\theta_{c_v}}\right)\ket{0}^{\otimes \left(\frac{\log{d}}{\kappa + 2}\right)}.
    \label{eq:token_position_conditions}
\end{eqnarray}
We note that the $\kappa +2$ term arises because the total number of qubits ($\log d$) is divided into registers of equal size for each embedding--two registers for tokens and positions, along with $\kappa$ additional registers. Here, we incorporate $\kappa = 1$ additional embeddings for physicochemical properties, detailed in Section \ref{sec:dataset}. Frequently, additional embeddings like molecular properties ($\mathbf{c}$) use the same embedding vector for each sequence element. For this case, Equation \ref{equation:classically_including_conditions} reduces to
\begin{eqnarray}
    \mathbf{z}_i = \mathbf{e}_i + \mathbf{p}_i + \sum^\kappa_{v=1}\mathbf{c}_v= \mathbf{e}_i + \mathbf{p}_i +  \mathbf{c}.
\end{eqnarray}

\begin{figure*}[!htbp]
    \centering
    \includegraphics[width=1.0\textwidth]{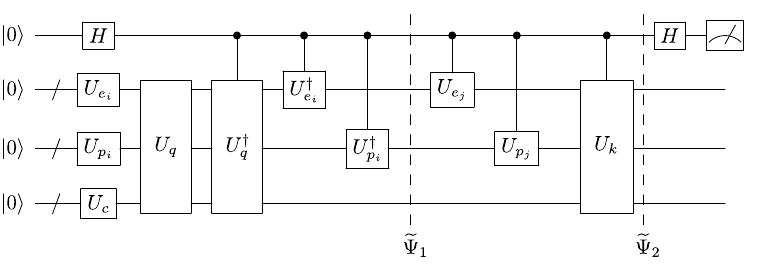}
    \caption{Quantum circuit in the hybrid quantum-classical self-attention mechanism. The circuit learns all embeddings, query and key representations, and produces a query-key dot product upon repeated measurement. $U_c$ is a set of angles representing the physicochemical property embeddings. Mathematical descriptions of $\widetilde{\Psi}_1$ and $\widetilde{\Psi}_2$ are found in Equations \ref{eq:psi_1_conditions} and \ref{eq:psi_2_conditions}, respectively.}
    \label{fig:circuit_with_conditions}
\end{figure*}

Likewise, the quantum circuit can be simplified when embeddings are uniform across the sequence. Instead of reversing the entire register to $\ket{0}^{\otimes \log d}$ under control, as in Equations \ref{eq:controlled_adjoint_circuit_1} and \ref{eq:controlled_adjoint_circuit_2}, subspaces with uniform embeddings remain unchanged, as they are identical across query-key pairs. This simplification yields the circuit in Figure \ref{fig:circuit_with_conditions}, where

\begin{eqnarray}
    \widetilde{\Psi}_1 = \frac{\ket{0}\ket{q_i} + \ket{1}\ket{0}^{\otimes \frac{2\log{d}}{3}}\ket{c}}{\sqrt{2}}
    \label{eq:psi_1_conditions}
\end{eqnarray}
\begin{eqnarray}
    \widetilde{\Psi}_2 = CU_{p_j}\left(CU_{e_j}\left(\widetilde{\Psi}_1\right)\right) =\nonumber \\
    \frac{\ket{0}\ket{q_i} + \ket{1}\ket{k_j}}{\sqrt{2}}.
    \label{eq:psi_2_conditions}
\end{eqnarray}
The term $\frac{2\log d}{3}$ arises because, as explained in Equation \ref{eq:token_position_conditions}, the quantum registers for each embedding are designed to have equal sizes. Consequently, with token, positional, and an additional $\kappa=1$ embedding representing molecular properties, there are three working quantum registers. Two of these three registers (the token and position quantum states) have been reversed under control.

\subsection{Quantum Gradient Calculation}
\subsubsection{Parameter Shift}
The inability to access intermediate quantum states significantly complicates traditional backpropagation via reverse-mode automatic differentiation for PQCs compared to classical methods \cite{abbas_quantum_2023}. The \textit{parameter shift rule} offers an approach for computing exact gradients of PQCs. The expectation value of an observable \( \hat{O} \) (e.g., a Pauli operator) is given by
\begin{eqnarray}
f(\theta) = \langle 0 | U^\dagger(\theta) \hat{O} U(\theta) | 0 \rangle
\end{eqnarray}

and the parameter shift rule computes its derivative with respect to \(\theta \) as follows
\begin{eqnarray}
\frac{\partial f(\theta)}{\partial \theta} = \frac{1}{2} \Big[ f\left(\theta + \frac{\pi}{2}\right) - f\left(\theta - \frac{\pi}{2}\right) \Big].
\end{eqnarray}

This approach computes the gradient by evaluating \( f\left(\theta\right) \) at shifted values \(\theta \pm \frac{\pi}{2}\), avoiding the need for finite differences but introducing an overhead of two circuit evaluations per parameter. As a result, its computational cost scales linearly with the number of parameters, rendering it impractical for large-scale QML models and motivating alternative methods.

\subsubsection{Simultaneous Perturbation Stochastic Approximation}
To address the parameter shift method’s linear scalability with parameter count, we employ approximate quantum gradient calculations via the simultaneous perturbation stochastic approximation (SPSA) algorithm \cite{spall_multivariate_1992}. SPSA seeks a parameter vector $\mathbf{x} \in \mathbb{R}^m$ that minimizes
\begin{eqnarray}
    \min_{\mathbf{x}} f(\mathbf{x}) = \min_{\mathbf{x}} \mathbb{E}_{\xi} \left[ F(\mathbf{x}, \xi) \right],
\end{eqnarray}

where $F: \mathbb{R}^m \to \mathbb{R}$ depends on the parameter vector $\mathbf{x} \in \mathbb{R}^m$ and a noise term $\xi$. SPSA approximates the gradient $\nabla f(\mathbf{x})$ as follows:

\begin{eqnarray}
    \nabla f(\textbf{x}) \approx \frac{f(\textbf{x}+\epsilon \cdot \bm{\Delta}) - f(\textbf{x}-\epsilon \cdot \bm{\Delta})}{2\epsilon\bm{\Delta}},
\end{eqnarray}
where $\mathbf{\Delta}$ is an $m$-dimensional vector with elements randomly chosen as $\pm 1$, and the perturbation step $\epsilon$ is set to $0.01$ in this work.

SPSA requires just two PQC evaluations regardless of parameter count, making it ideal for the NISQ era. Convergence analysis \cite{hoffmann_gradient_2022} indicates that larger batch sizes yield more stable updates by reducing gradient variance, aligning well with our training approach using a batch size of 256.

\subsection{Complexity Analysis}
\subsubsection{Computational Complexity}
The time complexity of the classical calculation of the attention matrix $\textbf{A}$ in Equation \ref{equation:classical_attention_matrix} is $\mathcal{O}(n^2d)$ stemming from the multiplication of $\textbf{Q}$ and $\textbf{K}^\top$ which are matrices of dimensions $n \times d$ and $d \times n$, respectively. Under the assumption of efficient preparation of quantum states, this modified-Hadamard test approach produces each inner product $\braket{q_i|k_j}$ in $\mathcal{O}(1)$ time compared to the classical $\mathcal{O}(d)$, leading to a complexity of $\mathcal{O}(n^2)$ for the population of $\textbf{A}$. In practice, we prepare states with $\mathcal{O}(\log{d})$ depth, bringing the overall practical complexity to $\mathcal{O}(n^2\log{d})$. It is important to note that preparing quantum states in $\mathcal{O}(\log{d})$ time may result in highly structured states. Specifically, low depth circuits are likely to produce quantum states with sparse amplitudes or pattern-like structures, which could make direct comparisons to classical algorithms that are designed for dense and unstructured data less equitable.

\subsubsection{Query Complexity}
The measurement of the ancilla qubit produces a Bernoulli random variable where the probability of measuring 1 is given by $p_0 = \frac{1+\text{Re}\braket{q_i|k_j}}{2}$. Thus, the Chernoff–Hoeffding theorem demonstrates $\text{Re}\braket{q_i|k_j}$ can be found in $\mathcal{O}\left(\frac{1}{\epsilon^2}\right)$ query complexity with additive error $\epsilon$. However, since $\text{Re}\braket{q_i|k_j} = 2p_0-1$ is obtained by measuring the ancilla qubit in the $\ket{0}$ state, quantum amplitude amplification can be performed to improve query complexity to $\mathcal{O}\left(\frac{1}{\epsilon}\right)$. To illustrate this, the final state before measurement as described in Equation \ref{eq:final_state} can be written equivalently:
\begin{eqnarray}
    \Psi_7 = \frac{\ket{0}}{2} \otimes \left(\ket{q_i}+\ket{k_j}\right) + \frac{\ket{1}}{2} \otimes \left(\ket{q_i}-\ket{k_j}\right) =\nonumber \\
    \sqrt{p_0}\ket{\psi_{good}} + \sqrt{1-p_0}\ket{\psi_{bad}}.
\end{eqnarray}
In this form, it is clear to see that a quadratic speed up in query complexity can be achieved using amplitude amplification with Grover operator 
\begin{eqnarray}
    G = R_{\Psi_7}R_{good}
\end{eqnarray}
where the reflector $R_{\Psi_7}$ is the unitary that prepares the entire Hadamard-test circuit $\Psi_7 = R_{\Psi_7}\ket{0}^{\otimes\left(\log{d}+1\right)}$, and the reflector $R_{good}$ is the Pauli-Z gate since the ancilla qubit is always in state $\ket{0}$ for $\ket{\psi_{good}}$. While we leave the implementation of amplitude amplification to future researchers, we note the overall practical complexity of the attention score calculation may be reduced to $\mathcal{O}\left(\frac{n^2\log{d}}{\epsilon}\right)$.

\section{Experiments}
\subsection{Dataset \& Features}\label{sec:dataset}
We employ the QM9 dataset \cite{ramakrishnan_quantum_2014}, a benchmark of 133,885 small organic molecules represented as SMILES strings \cite{weininger_smiles_1988}, as the basis for our study. The SMILES are canonicalized with RDKit \cite{greg_landrum_rdkitrdkit_2024} and duplicates were removed leaving 133,798 remaining datapoints. After preprocessing, the dataset was split into training and validation sets at a 20:1 ratio. From the dataset, we extracted nine key physicochemical features using RDKit: molecular weight (MW), number of hydrogen bond acceptors (HBA), number of hydrogen bond donors (HBD), number of rotatable bonds (nRot), number of rings (nRing), number of heteroatoms (nHet), topological polar surface area (TPSA), logP (partition coefficient), and the number of stereocenters (Stereo). To incorporate these descriptors as additional embeddings into our quantum-classical hybrid model, we transform them via a classical linear layer from 9 dimensions to $\dim{\theta_e} = \dim{\theta_p}$. We then scale the batch linearly between between $0$ and $\pi$ to produce $\theta_c$ for quantum circuit encoding. The SMILES strings are tokenized by breaking them into meaningful substructures, such as atoms, rings, branches, and bond types, and converting them into a sequence of discrete tokens that can be mapped to a high-dimensional embedding vector. The QM9 SMILES strings consist of 30 unique tokens, along with padding, start-of-sequence, and end-of-sequence tokens, resulting in a total vocabulary size of 33.

\subsection{Benchmarks \& Trainings}

We trained two models in this study: one learning SMILES strings using only text sequences, and another incorporating physicochemical embeddings for conditional molecular generation. For the sequence-only setup, we assigned 3 qubits to each of the token and positional registers. For the condition-based setup, we allocated 2 qubits to each of the token, positional, and physicochemical registers, maintaining 6 working qubits across both configurations.

We compared the quantum-classical model’s performance to that of fully classical models with equivalent architectures. All training setups employed one decoder layer and one attention head. The quantum model computed attention scores using 6 active qubits, producing a Hilbert space of dimension $2^6 = 64$, and thus we set the classical token and positional embedding vectors to 64 dimensions. We also trained a classical model with equal parameter counts, denoted Classical--eq, for further comparison. Since each PQC ansatz uses a single layer of $R_y$ gates, the number of learnable parameters per register matches the qubit count—3 for token and positional registers in sequence-only training ($\frac{\log 64}{2} = 3$, Equation \ref{equation:token_and_pos_unitaries}), and 2 for token, positional, and physicochemical registers in condition-based training ($\frac{\log 64}{3} = 2$, Equation \ref{eq:token_position_conditions}). To match the parameter count between the Quantum and Classical--eq models, the weight matrices $\mathbf{W}^Q$ and $\mathbf{W}^K$ (Equation \ref{eq:classical_Q_K_V}) have shapes $3 \times 2$ and $2 \times 3$ for sequence-only and condition-based setups, respectively, yielding 6 parameters each for query and key transformations. All models in this work use 64-dimensional embeddings for the value matrix $\mathbf{W}^V$. 
To summarize, the total number of parameters for the hybrid quantum-classical model (Quantum), fully classical model with an equal number of parameters (Classical--eq) model were 47,704 and 48,307 for the sequence and condition-based models, respectively. The fully classical model with an equivalent architecture to the Quantum model but with traditionally sized query-key weight matrices $\mathbf{W}^Q$ and $\mathbf{W}^K$ of shape $64 \times 64$ (denoted as Classical) has 55,713 and 56,535 parameters for the sequence and condition-based model, respectively. To fairly evaluate the performance of these models, we initialized shared parameters across all models with identical random values.

We implemented the machine learning components using PyTorch \cite{paszke_pytorch_2019}. All models underwent training for 20 epochs with the AdamW optimizer \cite{loshchilov_decoupled_2019}, set to a learning rate of 0.005 and weight decay of 0.1. We applied gradient clipping with a maximum norm of 1.0 per layer to stabilize gradients and used cross-entropy loss as the objective function. To support a batch size of 256, we conducted quantum circuit simulations with CUDA-Q \cite{The_CUDA-Q_development_team_CUDA-Q}, an open-source QPU-agnostic platform designed for accelerated quantum-classical supercomputing. All quantum simulations were performed using the state-vector simulator available in CUDA-Q. Training times for a single epoch on a CPU versus a single GPU are shown in Table \ref{tab:target_runtimes}, with a single GPU achieving a speedup of 1.34x over the CPU for the simulation with 7 qubits. The speedup here is moderate, as the size of the embedding is small and hence it does not fully saturate the GPU. Moreover, while the CPU time reflects pure computation, the GPU time includes data transfer overhead between host and device. For a larger embedding of a 13-qubit system, the GPU achieves a 5.75x speedup in runtime compared to the CPU. This gap is expected to widen further as we consider larger model sizes. 

Distributing simulations across four GPUs results in a 3.84x speedup for the 7-qubit system and a 3.96x speedup for the 13-qubit system, compared to using a single GPU. Consequently, we utilized four NVIDIA A100 GPUs on NERSC’s Perlmutter supercomputer to accelerate training with this large batch size.

\begin{table}[!h]
\caption{\label{tab:target_runtimes}%
Training time per epoch for the condition-based quantum model on the QM9 dataset using a batch size of 256. Values are extrapolated from the average runtime over four batch updates. All runs used an AMD EPYC 7763 CPU (64 cores, 128 threads, default thread settings for CUDA-Q and Pytorch) and NVIDIA A100 GPUs. For the 7-qubit model, GPU speedup over CPU is modest because the quantum circuits are relatively small, leading to limited GPU utilization and increased sensitivity to data transfer overhead between host and device. As system size increases, circuit complexity grows, allowing the GPU to more effectively leverage its parallel architecture. For the 13-qubit model, this results in a 5.75x speedup over CPU performance, and an additional 3.96x speedup when scaling across four GPUs.}

\begin{ruledtabular}
\begin{tabular}{ccc}
\textrm{Qubits}&
\textrm{Hardware}&
\textrm{Epoch Time (hrs)}\\

\colrule 
 & CPU &  41.28\\
7  & 1 GPU & 30.85\\
 & 4 GPUs & 8.03\\
\colrule
  & CPU &  424.34\\
13  & 1 GPU & 73.84\\
 & 4 GPUs & 18.65\\
\end{tabular}
\end{ruledtabular}
\end{table}
\

\FloatBarrier
\section{Results}
\subsection{Evaluation of SMILES String Learning \& Generation} \label{sec:smiles_generation_section}
\begin{figure}[h]
    \centering
    \subfloat[Sequence-only training loss]{
        \includegraphics[width=0.24\textwidth]{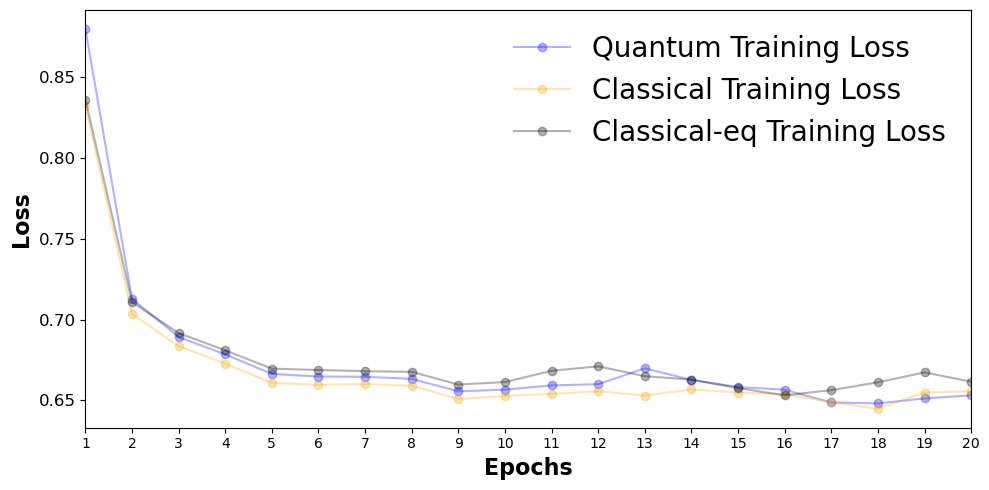}
        \label{fig:curve1}
    }
    \subfloat[Sequence-only validation loss]{
        \includegraphics[width=0.24\textwidth]{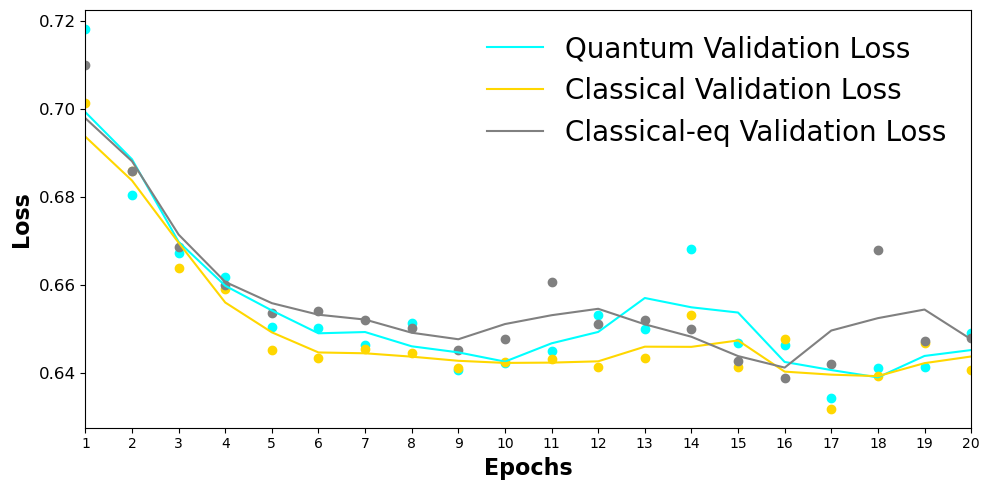}
        \label{fig:curve2}
    }

    \subfloat[Condition-based training loss]{
        \includegraphics[width=0.24\textwidth]{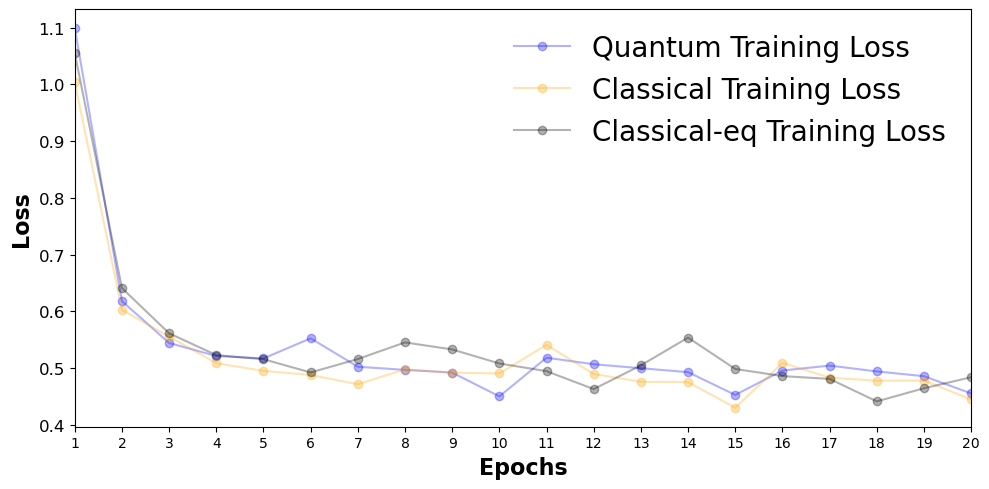}
        \label{fig:curve3}
    }
    \subfloat[Condition-based validation loss]{
        \includegraphics[width=0.24\textwidth]{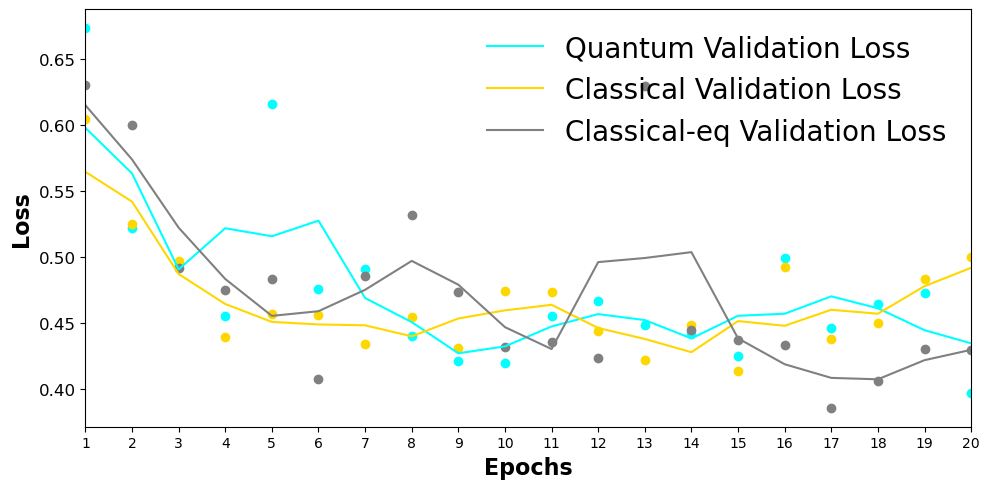}
        \label{fig:curve4}
    }

    \caption{Learning curves for the training and validation losses of the quantum-classical model (Quantum), the fully classical model with an equal number of parameters to the quantum-classical model (Classical -- eq), and the fully classical model with an equivalent architecture  but with traditionally sized weight matrices (Classical). To better illustrate the model's learning progress, the validation curves (\ref{fig:curve2}, \ref{fig:curve4}) display the 3-epoch moving average of the loss.}
    \label{fig:learning_curves}
\end{figure}

Training and validation loss curves for the sequence-based and condition-based models are presented in Figure \ref{fig:learning_curves}. For performance evaluation, we used the epoch with the lowest validation loss per model for inference, reporting next-token accuracy on the validation set as the ratio of correctly predicted tokens (highest output probability matching the true token) to total tokens. The validity, uniqueness, and novelty percentages from 100,000 inference queries to the trained models are reported. The results in Table \ref{tab:validity} for the condition-based model used the mean values of each physicochemical property in the training set to construct the property embedding vector during inference, as it produced the greatest rate of valid and unique SMILES (shown as \( V \times U \) in Table \ref{tab:validity}) across all models. The results using the mean, median, and mode of each property to guide the conditional generation were tested and are shown in Table \ref{tab:appendix_validity} in the appendix.

The decrease in \( V \times U \)—from 55.1–56.2\% for the sequence-only model to 18.5–20.3\% for the condition-based model arises from the trade-off between structural validity and diversity when optimizing for property constraints. The condition-based model samples molecules from a more narrow chemical space to satisfy both structural and physicochemical constraints. The benefit of incorporating property embeddings is reflected in the improved next-token prediction accuracy across all models, increasing from 61.6–62.4\% to 68.0–69.9\%. Across all metrics in both the sequence-only and condition-based trainings, the quantum and classical models exhibited comparable performances.

\begin{table*}[!ht]
\caption{\label{tab:validity}%
Performance of each model at the epoch with the lowest validation loss. Quantum denotes the quantum-classical hybrid model. Classical--eq denotes a fully classical model with an equal of number learnable parameters as the Quantum model. Classical denotes a fully classical model with an equivalent architecture as the Quantum model, but with traditionally sized weight matrices. Accuracy \% is the percentage of tokens correctly predicted. Validity \% is the percentage of generated sequences that create a valid \texttt{mol} structure in RDKit out of 100,000 queries to the trained model. The product of validity (V) and uniqueness (U) shows the percentage of model queries which result in unique compounds. Novelty \% is the percentage of valid and unique SMILES strings that do not appear in the training set. 
}
\begin{ruledtabular}
\begin{tabular}{cccccccc}
 &
\textrm{Model}&
\textrm{Loss}&
\textrm{Accuracy \%}&
\textrm{Validity \%}&
\textrm{Uniqueness \%}&
\textrm{V$\times$U \%}&
\textrm{Novelty \%}\\

\colrule
 & Quantum & 0.634 & 62.0 & 68.6 & 81.9 & 56.2 & 52.6\\
Sequence Only & Classical -- eq & 0.639 & 61.6 & 69.4 & 79.4 & 55.1 & 53.9\\
 & Classical& 0.632 & 62.4 & 72.5 & 81.2 & 55.9 & 52.0\\
\colrule
 & Quantum & 0.397 & 69.9 & 50.5 & 38.8 & 19.6 & 69.6\\
Property Embeddings & Classical -- eq & 0.386 & 68.3 & 50.7 & 40.0 & 20.3 & 70.4 \\
 & Classical & 0.414 & 68.0 & 38.5 & 48.0 & 18.5 & 71.2\\
\end{tabular}
\end{ruledtabular}
\end{table*}

\subsection{In-Distribution Modeling Performance}
Following SMILES generation (Section \ref{sec:smiles_generation_section}), we evaluated models trained with physicochemical embeddings for their ability to reproduce the training set’s property distribution. The first row of Table \ref{tab:conditions} presents the mean values of the nine molecular properties in the training set. To assess how well each model aligns with this distribution, we performed inference using the epoch with the lowest validation loss per model-- consistent with Section \ref{sec:smiles_generation_section}-- employing a physicochemical embedding vector based on these mean property values. After 100,000 queries, we computed the properties of all valid SMILES strings and reported their averages in the upper section of Table \ref{tab:conditions} for each model. Bold values in each column indicate the model generating valid molecules closest to the target mean. Results revealed comparable performance across models, with the quantum model producing molecules nearest to the target means for 3 of 9 properties: hydrogen bond acceptors (HBA), heteroatoms (nHet), and logP (octanol-water partition coefficient). The Classical--eq model excelled at generating molecules with on-target molecular weights (MW), while the Classical model outperformed others for the remaining five properties: hydrogen bond donors (HBD), rotatable bonds (nRot), rings (nRing), topological polar surface area (TPSA), and number of stereocenters (Stereo).

\subsection{Out-of-Distribution Modeling Performance}
The middle and bottom sections of Table \ref{tab:conditions} examined the models’ ability to generate molecules beyond the training distribution. For each experiment, we set the target for one property two standard deviations above and below the mean $(\mu \pm 2\sigma)$, imputing the other eight properties from the training data using the k-nearest neighbors (k-NN) method in scikit-learn \cite{pedregosa_scikit-learn_2011}.

For targets $2\sigma$ above the mean, the quantum model generated molecules closest to the targets for four properties: nRing, nHet, LogP, and Stereo. The Classical--eq model matched three properties (MW, HBD, TPSA), while the Classical model outperformed others on two (HBA, nRot). Below the mean by $2\sigma$, the quantum model excelled at HBD, nRot, TPSA, and LogP; the Classical--eq model at MW and nHet; and the Classical model at HBA and Stereo. Notably, all models generated molecules with zero rings equally well. To assess whether skewed distributions disproportionately affect any model, we repeated the experiment using median $\pm 1.5 $ IQR targets, where IQR is the interquartile range. Results in Table \ref{tab:appendix_conditions} confirm further that all models exhibit comparable performance in generating molecules beyond the training distribution.

\subsection{Comparison of Attention Maps}
\begin{figure}[!htbp]
    \centering
    \subfloat[{O=[N+]([O-])c1ccoc1}]{
        \makebox[0.24\textwidth][c]{\includegraphics[width=0.15\textwidth]{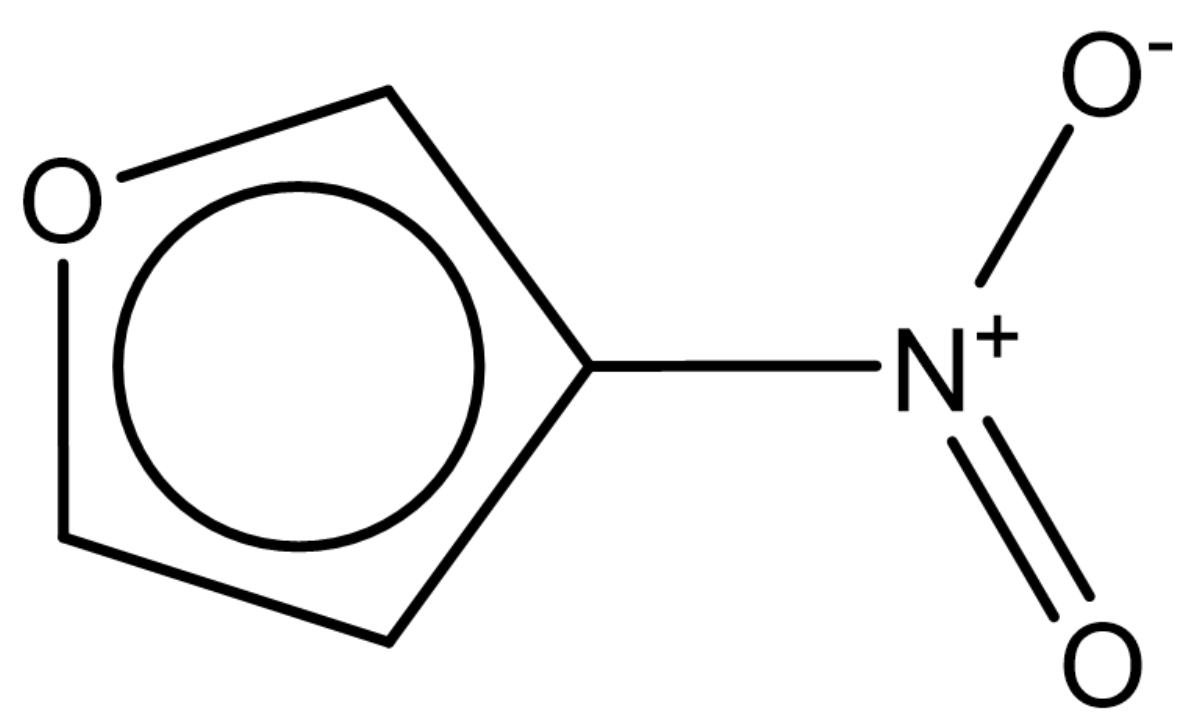}}
        \label{fig:O=[N+]([O-])c1ccoc1}
    }
    \subfloat[Quantum]{
        \includegraphics[width=0.24\textwidth]{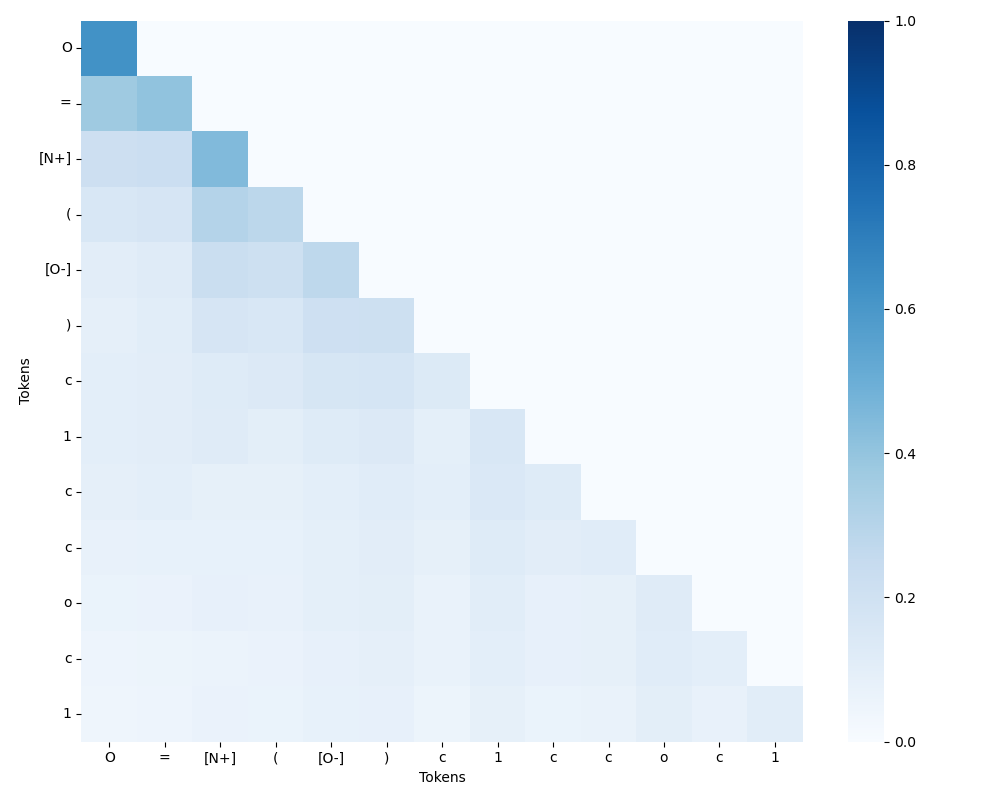}
        \label{fig:quantum_attention_map}
    }

    \subfloat[Classical -- eq]{
        \includegraphics[width=0.24\textwidth]{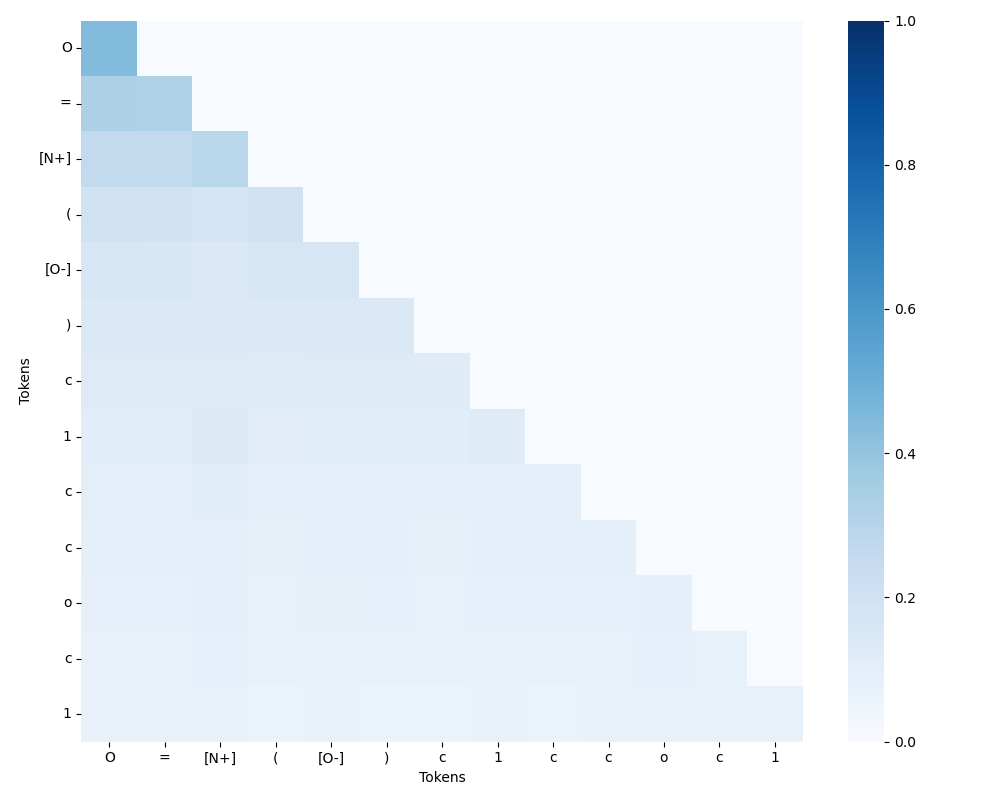}
        \label{fig:classical_reduced_attention_map}
    }
    \subfloat[Classical]{
        \includegraphics[width=0.24\textwidth]{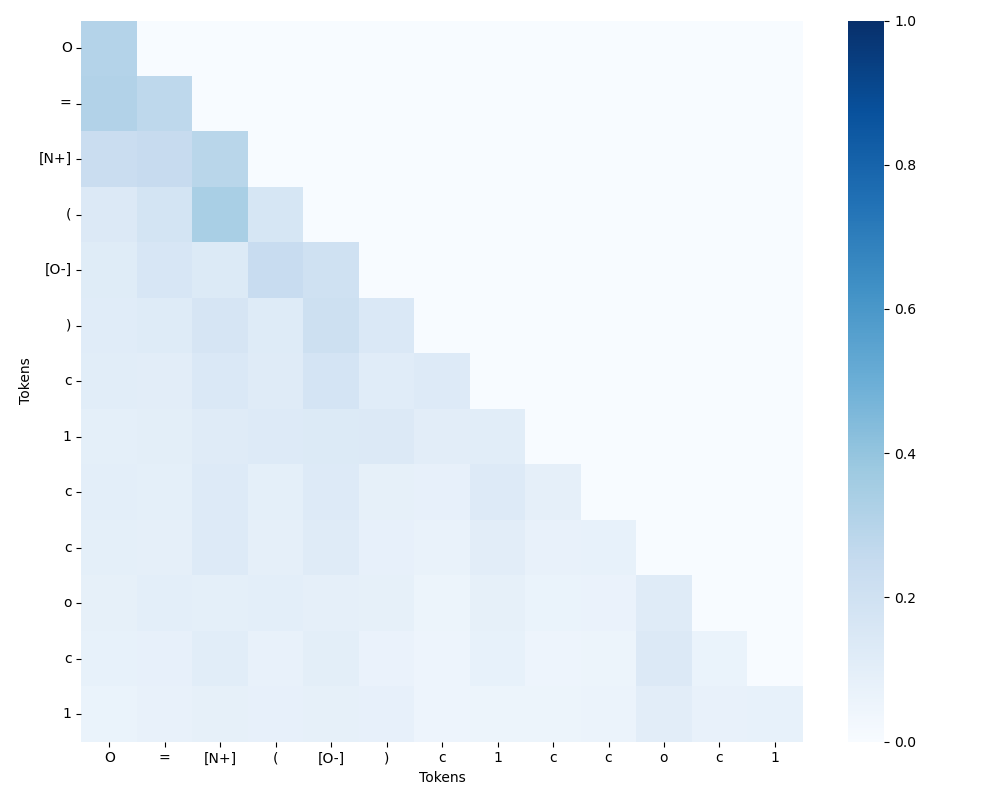}
        \label{fig:classical_attention_map}
    }

    \caption{Attention maps of O=[N+]([O-])c1ccoc1 for the quantum-classical hybrid model (Quantum) (\ref{fig:quantum_attention_map}), the fully classical model with an equal number of parameters (Classical--eq) (\ref{fig:classical_reduced_attention_map}), and the fully classical model with an equivalent architecture as the quantum-classical model but with traditionally sized weight matrices (Classical)(\ref{fig:classical_attention_map}). All attention maps were computed using the model parameters from the epoch with the lowest validation loss per condition-based model.}
    \label{fig:attention_maps}
\end{figure}

To visualize and qualitatively compare the features learned by the attention mechanisms, attention maps for an example molecule shown in Figure \ref{fig:O=[N+]([O-])c1ccoc1}, are presented in Figures \ref{fig:quantum_attention_map} -- \ref{fig:classical_attention_map}. While the aggregate quantitative performance of the models is similar, it is evident that they do not learn the same features to the same extent. This divergence in feature learning highlights the potential utility of hybrid quantum-classical self-attention mechanisms. Combining quantum and classical self-attention heads could enhance the extraction of a broader range of sequence features compared to relying solely on either. Such an approach could improve downstream task performance, an avenue for future research.

\begin{table*}[!ht]
\caption{\label{tab:conditions}%
Conditional generation results. The top section demonstrates how well each model is able to generate molecules targeting the mean values of each property from the training data. The middle and lower sections indicate a target that is above and below the mean value for each property by 2 $\times$ the standard deviation ($\sigma$), respectively. For each model, the average value for that property of all valid generated molecules are shown. In the middle and lower sections, each numerical entry represents the result from an inference experiment where only that property was specified and the remaining 8 properties were imputed from the training data with k-nearest neighbors. Bold values indicate which model generated molecules closer to the target value. Quantum indicates the quantum-classical hybrid model, Classical -- eq denotes the fully classical model with an equal number parameters as the Quantum model, Classical denotes the fully classical model with an equivalent architecture to the Quantum model, but with traditionally sized weight matrices. All inferences were performed with the epoch that possessed the lowest validation loss for each model.
}
\begin{ruledtabular}
\begin{tabular}{cccccccccc}
 &
\textrm{MW}&
\textrm{HBA}&
\textrm{HBD}&
\textrm{nRot}&
\textrm{nRing}&
\textrm{nHet}&
\textrm{TPSA}&
\textrm{logP}&
\textrm{Stereo}\\

\colrule

Mean & 122.77	&2.23	&0.83	&0.92	&1.74	&2.47	&37.16	&0.30	&1.71 \\

Quantum & 125.13 & \textbf{2.31} & 0.55 & 0.63 & 2.05 & \textbf{2.37} & 32.58 & \textbf{0.30} & 2.11 \\
Classical -- eq & \textbf{123.42} & 2.15 & 0.55 & 0.55 & 2.03 & 2.21 & 31.74 & 0.46 & 1.98 \\
Classical & 128.04 & 2.15 & \textbf{1.00} & \textbf{0.79} & \textbf{1.67} & 2.19 & \textbf{34.53} & 0.52 & \textbf{1.81} \\

\colrule
Mean + ($2 \times \sigma$) & 137.88 & 4.34 & 2.50 & 3.10 & 4.16 & 4.84 & 79.67 & 2.30 & 4.77 \\

Quantum & 135.73 & 3.95 & 2.79 & 2.71 & \textbf{3.74} & \textbf{4.82} & 81.61 & \textbf{2.24} & \textbf{4.46} \\
Classical -- eq & \textbf{137.12} & 3.98 & \textbf{2.75} & 2.52 & 3.62 & 4.76 & \textbf{78.03} & 2.36 & 4.34 \\
Classical & 136.09 & \textbf{4.05} & 2.90 & \textbf{2.91} & 3.53 & 4.78 & 83.44 & 2.37 & 4.27 \\

\colrule
Mean - ($2 \times \sigma$) & 107.66 & 0.12 & 0.00 & 0.00 & 0.00 & 0.10 & 0.00 & -1.71 & 0.00 \\
 
Quantum & 112.83 & 0.63 & \textbf{0.15} & \textbf{0.11} & 0.00 & 0.05 & \textbf{0.55} & \textbf{-1.74} & 0.41 \\
Classical -- eq & \textbf{112.12} & 0.58 & 0.19 & 0.11 & 0.00 & \textbf{0.06} & 0.67 & -1.49 & 0.26 \\
Classical & 113.33 & \textbf{0.30} & 0.33 & 0.11 & 0.00 & 0.06 & 0.66 & -1.46 & \textbf{0.15} \\

\end{tabular}
\end{ruledtabular}
\end{table*}

\section{Limitations \& Conclusions}
Our primary contribution is demonstrating that quantum states and learnable unitary evolutions can replace classical self-attention components in a generative model, achieving a NISQ-friendly solution that maintains performance parity with fully classical models. Additionally, we introduce a novel method for incorporating positional encodings to enhance the model's ability to learn sequence-based information, alongside the integration of supplementary embeddings, such as molecular properties. This approach enabled targeted molecular generation, producing molecules with desired properties, thus demonstrating the potential of hybrid quantum-classical architectures for generative tasks. Notably, we achieved comparable performance between our quantum and classical baselines while training with the Simultaneous Perturbation Stochastic Approximation algorithm. Since parameter-shift gradients are computationally expensive, and backpropagation remains a common criticism of quantum neural networks, demonstrating competitive performance using SPSA is a promising result.

Despite these advances, our method has limitations to consider. The primary bottleneck in a self-attention mechanism's complexity is its quadratic scaling with sequence length, $\mathcal{O}(n^2d)$. Our proposed method reduces the attention matrix computation time complexity to $\mathcal{O}(n^2\log{d})$ but fails to address the dominant $n^2$ quadratic scaling term. Additionally, the complexity to multiply the attention matrix and value matrix still scales $\mathcal{O}(n^2d)$. While prior quantum transformer and self-attention formulations suggest further reductions are theoretically possible \cite{liao_gpt_2024, guo_quantum_2024}, they demand quantum resources impractical for the NISQ era, reinforcing our focus on NISQ compatibility over exhaustive complexity optimization. We defer exploring additional attention heads, decoder layers, and more expressive ansatzes to future research. Remarkably, both the Quantum and Classical--eq models, with as few as two learnable parameters for tokens and positions, effectively learn SMILES strings. We hope these findings spur further development of practical, NISQ-ready designs that balance efficiency and performance for generative modeling.

\section{Data Availability}
The datasets and code to reproduce the figures and results from this work are available at \href{https://github.com/anthonysmaldone/Quantum-Transformer}{https://github.com/anthonysmaldone/Quantum-Transformer}.

\begin{acknowledgments}
The authors acknowledge support from the National Science Foundation Engines Development Award: Advancing Quantum Technologies (CT) under Award Number 2302908. AMS and GWK acknowledge the financial support from the National Science Foundation Graduate Research Fellowship under Award Number DGE-2139841. VSB also acknowledges partial support from the National Science Foundation Center for Quantum Dynamics on Modular Quantum Devices (CQD-MQD) under Award Number 2124511, as well as computational resources provided by the National Energy Research Scientific Computing Center (NERSC) under the DOE Mission Science award ERCAP0031864.
\end{acknowledgments}
\bibliography{zotero_bibtex_truncated, additional_references}

\begin{thebibliography}{33}%
\makeatletter
\providecommand \@ifxundefined [1]{%
 \@ifx{#1\undefined}
}%
\providecommand \@ifnum [1]{%
 \ifnum #1\expandafter \@firstoftwo
 \else \expandafter \@secondoftwo
 \fi
}%
\providecommand \@ifx [1]{%
 \ifx #1\expandafter \@firstoftwo
 \else \expandafter \@secondoftwo
 \fi
}%
\providecommand \natexlab [1]{#1}%
\providecommand \enquote  [1]{``#1''}%
\providecommand \bibnamefont  [1]{#1}%
\providecommand \bibfnamefont [1]{#1}%
\providecommand \citenamefont [1]{#1}%
\providecommand \href@noop [0]{\@secondoftwo}%
\providecommand \href [0]{\begingroup \@sanitize@url \@href}%
\providecommand \@href[1]{\@@startlink{#1}\@@href}%
\providecommand \@@href[1]{\endgroup#1\@@endlink}%
\providecommand \@sanitize@url [0]{\catcode `\\12\catcode `\$12\catcode `\&12\catcode `\#12\catcode `\^12\catcode `\_12\catcode `\%12\relax}%
\providecommand \@@startlink[1]{}%
\providecommand \@@endlink[0]{}%
\providecommand \url  [0]{\begingroup\@sanitize@url \@url }%
\providecommand \@url [1]{\endgroup\@href {#1}{\urlprefix }}%
\providecommand \urlprefix  [0]{URL }%
\providecommand \Eprint [0]{\href }%
\providecommand \doibase [0]{https://doi.org/}%
\providecommand \selectlanguage [0]{\@gobble}%
\providecommand \bibinfo  [0]{\@secondoftwo}%
\providecommand \bibfield  [0]{\@secondoftwo}%
\providecommand \translation [1]{[#1]}%
\providecommand \BibitemOpen [0]{}%
\providecommand \bibitemStop [0]{}%
\providecommand \bibitemNoStop [0]{.\EOS\space}%
\providecommand \EOS [0]{\spacefactor3000\relax}%
\providecommand \BibitemShut  [1]{\csname bibitem#1\endcsname}%
\let\auto@bib@innerbib\@empty
\bibitem [{\citenamefont {Vaswani}\ \emph {et~al.}(2017)\citenamefont {Vaswani}, \citenamefont {Shazeer}, \citenamefont {Parmar}, \citenamefont {Uszkoreit}, \citenamefont {Jones}, \citenamefont {Gomez}, \citenamefont {Kaiser},\ and\ \citenamefont {Polosukhin}}]{vaswani_attention_2017}%
  \BibitemOpen
  \bibfield  {author} {\bibinfo {author} {\bibfnamefont {A.}~\bibnamefont {Vaswani}}, \bibinfo {author} {\bibfnamefont {N.}~\bibnamefont {Shazeer}}, \bibinfo {author} {\bibfnamefont {N.}~\bibnamefont {Parmar}}, \bibinfo {author} {\bibfnamefont {J.}~\bibnamefont {Uszkoreit}}, \bibinfo {author} {\bibfnamefont {L.}~\bibnamefont {Jones}}, \bibinfo {author} {\bibfnamefont {A.~N.}\ \bibnamefont {Gomez}}, \bibinfo {author} {\bibfnamefont {L.}~\bibnamefont {Kaiser}},\ and\ \bibinfo {author} {\bibfnamefont {I.}~\bibnamefont {Polosukhin}},\ }in\ \href {https://papers.nips.cc/paper_files/paper/2017/hash/3f5ee243547dee91fbd053c1c4a845aa-Abstract.html} {\emph {\bibinfo {booktitle} {Advances in {Neural} {Information} {Processing} {Systems}}}},\ Vol.~\bibinfo {volume} {30}\ (\bibinfo  {publisher} {Curran Associates, Inc.},\ \bibinfo {year} {2017})\BibitemShut {NoStop}%
\bibitem [{\citenamefont {{OpenAI}}\ \emph {et~al.}(2024)\citenamefont {{OpenAI}}, \citenamefont {Achiam}, \citenamefont {Adler}, \citenamefont {Agarwal}, \citenamefont {Ahmad}, \citenamefont {Akkaya}, \citenamefont {Aleman}, \citenamefont {Almeida}, \citenamefont {Altenschmidt}, \citenamefont {Altman}, \citenamefont {Anadkat}, \citenamefont {Avila}, \citenamefont {Babuschkin}, \citenamefont {Balaji}, \citenamefont {Balcom},\ and\ \citenamefont {{et al.}}}]{openai_gpt-4_2024}%
  \BibitemOpen
  \bibfield  {author} {\bibinfo {author} {\bibnamefont {{OpenAI}}}, \bibinfo {author} {\bibfnamefont {J.}~\bibnamefont {Achiam}}, \bibinfo {author} {\bibfnamefont {S.}~\bibnamefont {Adler}}, \bibinfo {author} {\bibfnamefont {S.}~\bibnamefont {Agarwal}}, \bibinfo {author} {\bibfnamefont {L.}~\bibnamefont {Ahmad}}, \bibinfo {author} {\bibfnamefont {I.}~\bibnamefont {Akkaya}}, \bibinfo {author} {\bibfnamefont {F.~L.}\ \bibnamefont {Aleman}}, \bibinfo {author} {\bibfnamefont {D.}~\bibnamefont {Almeida}}, \bibinfo {author} {\bibfnamefont {J.}~\bibnamefont {Altenschmidt}}, \bibinfo {author} {\bibfnamefont {S.}~\bibnamefont {Altman}}, \bibinfo {author} {\bibfnamefont {S.}~\bibnamefont {Anadkat}}, \bibinfo {author} {\bibfnamefont {R.}~\bibnamefont {Avila}}, \bibinfo {author} {\bibfnamefont {I.}~\bibnamefont {Babuschkin}}, \bibinfo {author} {\bibfnamefont {S.}~\bibnamefont {Balaji}}, \bibinfo {author} {\bibfnamefont {V.}~\bibnamefont {Balcom}},\ and\ \bibinfo {author} {\bibnamefont {{et al.}}},\ }\href
  {https://doi.org/10.48550/arXiv.2303.08774} {\bibinfo {title} {{GPT}-4 {Technical} {Report}}} (\bibinfo {year} {2024}),\ \bibinfo {note} {arXiv:2303.08774 [cs]}\BibitemShut {NoStop}%
\bibitem [{\citenamefont {Dosovitskiy}\ \emph {et~al.}(2021)\citenamefont {Dosovitskiy}, \citenamefont {Beyer}, \citenamefont {Kolesnikov}, \citenamefont {Weissenborn}, \citenamefont {Zhai}, \citenamefont {Unterthiner}, \citenamefont {Dehghani}, \citenamefont {Minderer}, \citenamefont {Heigold}, \citenamefont {Gelly}, \citenamefont {Uszkoreit},\ and\ \citenamefont {Houlsby}}]{dosovitskiy_image_2021}%
  \BibitemOpen
  \bibfield  {author} {\bibinfo {author} {\bibfnamefont {A.}~\bibnamefont {Dosovitskiy}}, \bibinfo {author} {\bibfnamefont {L.}~\bibnamefont {Beyer}}, \bibinfo {author} {\bibfnamefont {A.}~\bibnamefont {Kolesnikov}}, \bibinfo {author} {\bibfnamefont {D.}~\bibnamefont {Weissenborn}}, \bibinfo {author} {\bibfnamefont {X.}~\bibnamefont {Zhai}}, \bibinfo {author} {\bibfnamefont {T.}~\bibnamefont {Unterthiner}}, \bibinfo {author} {\bibfnamefont {M.}~\bibnamefont {Dehghani}}, \bibinfo {author} {\bibfnamefont {M.}~\bibnamefont {Minderer}}, \bibinfo {author} {\bibfnamefont {G.}~\bibnamefont {Heigold}}, \bibinfo {author} {\bibfnamefont {S.}~\bibnamefont {Gelly}}, \bibinfo {author} {\bibfnamefont {J.}~\bibnamefont {Uszkoreit}},\ and\ \bibinfo {author} {\bibfnamefont {N.}~\bibnamefont {Houlsby}},\ }\href {https://doi.org/10.48550/arXiv.2010.11929} {\bibinfo {title} {An {Image} is {Worth} 16x16 {Words}: {Transformers} for {Image} {Recognition} at {Scale}}} (\bibinfo {year} {2021}),\ \bibinfo {note} {arXiv:2010.11929
  [cs]}\BibitemShut {NoStop}%
\bibitem [{\citenamefont {Carion}\ \emph {et~al.}(2020)\citenamefont {Carion}, \citenamefont {Massa}, \citenamefont {Synnaeve}, \citenamefont {Usunier}, \citenamefont {Kirillov},\ and\ \citenamefont {Zagoruyko}}]{carion_end--end_2020}%
  \BibitemOpen
  \bibfield  {author} {\bibinfo {author} {\bibfnamefont {N.}~\bibnamefont {Carion}}, \bibinfo {author} {\bibfnamefont {F.}~\bibnamefont {Massa}}, \bibinfo {author} {\bibfnamefont {G.}~\bibnamefont {Synnaeve}}, \bibinfo {author} {\bibfnamefont {N.}~\bibnamefont {Usunier}}, \bibinfo {author} {\bibfnamefont {A.}~\bibnamefont {Kirillov}},\ and\ \bibinfo {author} {\bibfnamefont {S.}~\bibnamefont {Zagoruyko}},\ }\href {https://doi.org/10.48550/arXiv.2005.12872} {\bibinfo {title} {End-to-{End} {Object} {Detection} with {Transformers}}} (\bibinfo {year} {2020}),\ \bibinfo {note} {arXiv:2005.12872 [cs]}\BibitemShut {NoStop}%
\bibitem [{\citenamefont {Jumper}\ \emph {et~al.}(2021)\citenamefont {Jumper}, \citenamefont {Evans}, \citenamefont {Pritzel}, \citenamefont {Green}, \citenamefont {Figurnov}, \citenamefont {Ronneberger}, \citenamefont {Tunyasuvunakool}, \citenamefont {Bates}, \citenamefont {Žídek}, \citenamefont {Potapenko}, \citenamefont {Bridgland}, \citenamefont {Meyer}, \citenamefont {Kohl}, \citenamefont {Ballard}, \citenamefont {Cowie},\ and\ \citenamefont {{et al.}}}]{jumper_highly_2021}%
  \BibitemOpen
  \bibfield  {author} {\bibinfo {author} {\bibfnamefont {J.}~\bibnamefont {Jumper}}, \bibinfo {author} {\bibfnamefont {R.}~\bibnamefont {Evans}}, \bibinfo {author} {\bibfnamefont {A.}~\bibnamefont {Pritzel}}, \bibinfo {author} {\bibfnamefont {T.}~\bibnamefont {Green}}, \bibinfo {author} {\bibfnamefont {M.}~\bibnamefont {Figurnov}}, \bibinfo {author} {\bibfnamefont {O.}~\bibnamefont {Ronneberger}}, \bibinfo {author} {\bibfnamefont {K.}~\bibnamefont {Tunyasuvunakool}}, \bibinfo {author} {\bibfnamefont {R.}~\bibnamefont {Bates}}, \bibinfo {author} {\bibfnamefont {A.}~\bibnamefont {Žídek}}, \bibinfo {author} {\bibfnamefont {A.}~\bibnamefont {Potapenko}}, \bibinfo {author} {\bibfnamefont {A.}~\bibnamefont {Bridgland}}, \bibinfo {author} {\bibfnamefont {C.}~\bibnamefont {Meyer}}, \bibinfo {author} {\bibfnamefont {S.~A.~A.}\ \bibnamefont {Kohl}}, \bibinfo {author} {\bibfnamefont {A.~J.}\ \bibnamefont {Ballard}}, \bibinfo {author} {\bibfnamefont {A.}~\bibnamefont {Cowie}},\ and\ \bibinfo {author} {\bibnamefont {{et
  al.}}},\ }\href {https://doi.org/10.1038/s41586-021-03819-2} {\bibfield  {journal} {\bibinfo  {journal} {Nature}\ }\textbf {\bibinfo {volume} {596}},\ \bibinfo {pages} {583} (\bibinfo {year} {2021})}\BibitemShut {NoStop}%
\bibitem [{\citenamefont {Abramson}\ \emph {et~al.}(2024)\citenamefont {Abramson}, \citenamefont {Adler}, \citenamefont {Dunger}, \citenamefont {Evans}, \citenamefont {Green}, \citenamefont {Pritzel}, \citenamefont {Ronneberger}, \citenamefont {Willmore}, \citenamefont {Ballard}, \citenamefont {Bambrick}, \citenamefont {Bodenstein}, \citenamefont {Evans}, \citenamefont {Hung}, \citenamefont {O’Neill}, \citenamefont {Reiman},\ and\ \citenamefont {{et al.}}}]{abramson_accurate_2024}%
  \BibitemOpen
  \bibfield  {author} {\bibinfo {author} {\bibfnamefont {J.}~\bibnamefont {Abramson}}, \bibinfo {author} {\bibfnamefont {J.}~\bibnamefont {Adler}}, \bibinfo {author} {\bibfnamefont {J.}~\bibnamefont {Dunger}}, \bibinfo {author} {\bibfnamefont {R.}~\bibnamefont {Evans}}, \bibinfo {author} {\bibfnamefont {T.}~\bibnamefont {Green}}, \bibinfo {author} {\bibfnamefont {A.}~\bibnamefont {Pritzel}}, \bibinfo {author} {\bibfnamefont {O.}~\bibnamefont {Ronneberger}}, \bibinfo {author} {\bibfnamefont {L.}~\bibnamefont {Willmore}}, \bibinfo {author} {\bibfnamefont {A.~J.}\ \bibnamefont {Ballard}}, \bibinfo {author} {\bibfnamefont {J.}~\bibnamefont {Bambrick}}, \bibinfo {author} {\bibfnamefont {S.~W.}\ \bibnamefont {Bodenstein}}, \bibinfo {author} {\bibfnamefont {D.~A.}\ \bibnamefont {Evans}}, \bibinfo {author} {\bibfnamefont {C.-C.}\ \bibnamefont {Hung}}, \bibinfo {author} {\bibfnamefont {M.}~\bibnamefont {O’Neill}}, \bibinfo {author} {\bibfnamefont {D.}~\bibnamefont {Reiman}},\ and\ \bibinfo {author} {\bibnamefont
  {{et al.}}},\ }\href {https://doi.org/10.1038/s41586-024-07487-w} {\bibfield  {journal} {\bibinfo  {journal} {Nature}\ }\textbf {\bibinfo {volume} {630}},\ \bibinfo {pages} {493} (\bibinfo {year} {2024})}\BibitemShut {NoStop}%
\bibitem [{\citenamefont {Biamonte}\ \emph {et~al.}(2017)\citenamefont {Biamonte}, \citenamefont {Wittek}, \citenamefont {Pancotti}, \citenamefont {Rebentrost}, \citenamefont {Wiebe},\ and\ \citenamefont {Lloyd}}]{biamonte_quantum_2017}%
  \BibitemOpen
  \bibfield  {author} {\bibinfo {author} {\bibfnamefont {J.}~\bibnamefont {Biamonte}}, \bibinfo {author} {\bibfnamefont {P.}~\bibnamefont {Wittek}}, \bibinfo {author} {\bibfnamefont {N.}~\bibnamefont {Pancotti}}, \bibinfo {author} {\bibfnamefont {P.}~\bibnamefont {Rebentrost}}, \bibinfo {author} {\bibfnamefont {N.}~\bibnamefont {Wiebe}},\ and\ \bibinfo {author} {\bibfnamefont {S.}~\bibnamefont {Lloyd}},\ }\href {https://doi.org/10.1038/nature23474} {\bibfield  {journal} {\bibinfo  {journal} {Nature}\ }\textbf {\bibinfo {volume} {549}},\ \bibinfo {pages} {195} (\bibinfo {year} {2017})},\ \bibinfo {note} {arXiv:1611.09347 [cond-mat, physics:quant-ph, stat]}\BibitemShut {NoStop}%
\bibitem [{\citenamefont {Wiebe}\ \emph {et~al.}(2012)\citenamefont {Wiebe}, \citenamefont {Braun},\ and\ \citenamefont {Lloyd}}]{wiebe_quantum_2012}%
  \BibitemOpen
  \bibfield  {author} {\bibinfo {author} {\bibfnamefont {N.}~\bibnamefont {Wiebe}}, \bibinfo {author} {\bibfnamefont {D.}~\bibnamefont {Braun}},\ and\ \bibinfo {author} {\bibfnamefont {S.}~\bibnamefont {Lloyd}},\ }\href {https://doi.org/10.1103/PhysRevLett.109.050505} {\bibfield  {journal} {\bibinfo  {journal} {Physical Review Letters}\ }\textbf {\bibinfo {volume} {109}},\ \bibinfo {pages} {050505} (\bibinfo {year} {2012})}\BibitemShut {NoStop}%
\bibitem [{\citenamefont {Lloyd}\ \emph {et~al.}(2014)\citenamefont {Lloyd}, \citenamefont {Mohseni},\ and\ \citenamefont {Rebentrost}}]{lloyd_quantum_2014}%
  \BibitemOpen
  \bibfield  {author} {\bibinfo {author} {\bibfnamefont {S.}~\bibnamefont {Lloyd}}, \bibinfo {author} {\bibfnamefont {M.}~\bibnamefont {Mohseni}},\ and\ \bibinfo {author} {\bibfnamefont {P.}~\bibnamefont {Rebentrost}},\ }\href {https://doi.org/10.1038/nphys3029} {\bibfield  {journal} {\bibinfo  {journal} {Nature Physics}\ }\textbf {\bibinfo {volume} {10}},\ \bibinfo {pages} {631} (\bibinfo {year} {2014})}\BibitemShut {NoStop}%
\bibitem [{\citenamefont {Rebentrost}\ \emph {et~al.}(2014)\citenamefont {Rebentrost}, \citenamefont {Mohseni},\ and\ \citenamefont {Lloyd}}]{rebentrost_quantum_2014}%
  \BibitemOpen
  \bibfield  {author} {\bibinfo {author} {\bibfnamefont {P.}~\bibnamefont {Rebentrost}}, \bibinfo {author} {\bibfnamefont {M.}~\bibnamefont {Mohseni}},\ and\ \bibinfo {author} {\bibfnamefont {S.}~\bibnamefont {Lloyd}},\ }\href {https://doi.org/10.1103/PhysRevLett.113.130503} {\bibfield  {journal} {\bibinfo  {journal} {Physical Review Letters}\ }\textbf {\bibinfo {volume} {113}},\ \bibinfo {pages} {130503} (\bibinfo {year} {2014})}\BibitemShut {NoStop}%
\bibitem [{\citenamefont {Cong}\ \emph {et~al.}(2019)\citenamefont {Cong}, \citenamefont {Choi},\ and\ \citenamefont {Lukin}}]{cong_quantum_2019}%
  \BibitemOpen
  \bibfield  {author} {\bibinfo {author} {\bibfnamefont {I.}~\bibnamefont {Cong}}, \bibinfo {author} {\bibfnamefont {S.}~\bibnamefont {Choi}},\ and\ \bibinfo {author} {\bibfnamefont {M.~D.}\ \bibnamefont {Lukin}},\ }\href {https://doi.org/10.1038/s41567-019-0648-8} {\bibfield  {journal} {\bibinfo  {journal} {Nature Physics}\ }\textbf {\bibinfo {volume} {15}},\ \bibinfo {pages} {1273} (\bibinfo {year} {2019})}\BibitemShut {NoStop}%
\bibitem [{\citenamefont {Zoufal}\ \emph {et~al.}(2019)\citenamefont {Zoufal}, \citenamefont {Lucchi},\ and\ \citenamefont {Woerner}}]{zoufal_quantum_2019}%
  \BibitemOpen
  \bibfield  {author} {\bibinfo {author} {\bibfnamefont {C.}~\bibnamefont {Zoufal}}, \bibinfo {author} {\bibfnamefont {A.}~\bibnamefont {Lucchi}},\ and\ \bibinfo {author} {\bibfnamefont {S.}~\bibnamefont {Woerner}},\ }\href {https://doi.org/10.1038/s41534-019-0223-2} {\bibfield  {journal} {\bibinfo  {journal} {npj Quantum Information}\ }\textbf {\bibinfo {volume} {5}},\ \bibinfo {pages} {1} (\bibinfo {year} {2019})}\BibitemShut {NoStop}%
\bibitem [{\citenamefont {Loshchilov}\ \emph {et~al.}(2024)\citenamefont {Loshchilov}, \citenamefont {Hsieh}, \citenamefont {Sun},\ and\ \citenamefont {Ginsburg}}]{loshchilov_ngpt_2024}%
  \BibitemOpen
  \bibfield  {author} {\bibinfo {author} {\bibfnamefont {I.}~\bibnamefont {Loshchilov}}, \bibinfo {author} {\bibfnamefont {C.-P.}\ \bibnamefont {Hsieh}}, \bibinfo {author} {\bibfnamefont {S.}~\bibnamefont {Sun}},\ and\ \bibinfo {author} {\bibfnamefont {B.}~\bibnamefont {Ginsburg}},\ }\href {https://doi.org/10.48550/arXiv.2410.01131} {\bibinfo {title} {{nGPT}: {Normalized} {Transformer} with {Representation} {Learning} on the {Hypersphere}}} (\bibinfo {year} {2024}),\ \bibinfo {note} {arXiv:2410.01131 [cs]}\BibitemShut {NoStop}%
\bibitem [{\citenamefont {Li}\ \emph {et~al.}(2023)\citenamefont {Li}, \citenamefont {Zhao},\ and\ \citenamefont {Wang}}]{li_quantum_2023}%
  \BibitemOpen
  \bibfield  {author} {\bibinfo {author} {\bibfnamefont {G.}~\bibnamefont {Li}}, \bibinfo {author} {\bibfnamefont {X.}~\bibnamefont {Zhao}},\ and\ \bibinfo {author} {\bibfnamefont {X.}~\bibnamefont {Wang}},\ }\href {https://doi.org/10.48550/arXiv.2205.05625} {\bibinfo {title} {Quantum {Self}-{Attention} {Neural} {Networks} for {Text} {Classification}}} (\bibinfo {year} {2023}),\ \bibinfo {note} {arXiv:2205.05625 [quant-ph]}\BibitemShut {NoStop}%
\bibitem [{\citenamefont {Khatri}\ \emph {et~al.}(2024)\citenamefont {Khatri}, \citenamefont {Matos}, \citenamefont {Coopmans},\ and\ \citenamefont {Clark}}]{khatri_quixer_2024}%
  \BibitemOpen
  \bibfield  {author} {\bibinfo {author} {\bibfnamefont {N.}~\bibnamefont {Khatri}}, \bibinfo {author} {\bibfnamefont {G.}~\bibnamefont {Matos}}, \bibinfo {author} {\bibfnamefont {L.}~\bibnamefont {Coopmans}},\ and\ \bibinfo {author} {\bibfnamefont {S.}~\bibnamefont {Clark}},\ }\href {https://doi.org/10.48550/arXiv.2406.04305} {\bibinfo {title} {Quixer: {A} {Quantum} {Transformer} {Model}}} (\bibinfo {year} {2024}),\ \bibinfo {note} {arXiv:2406.04305 [quant-ph]}\BibitemShut {NoStop}%
\bibitem [{\citenamefont {Zheng}\ \emph {et~al.}(2023)\citenamefont {Zheng}, \citenamefont {Gao},\ and\ \citenamefont {Miao}}]{zheng_design_2023}%
  \BibitemOpen
  \bibfield  {author} {\bibinfo {author} {\bibfnamefont {J.}~\bibnamefont {Zheng}}, \bibinfo {author} {\bibfnamefont {Q.}~\bibnamefont {Gao}},\ and\ \bibinfo {author} {\bibfnamefont {Z.}~\bibnamefont {Miao}},\ }in\ \href {https://doi.org/10.1109/SMC53992.2023.10393989} {\emph {\bibinfo {booktitle} {2023 {IEEE} {International} {Conference} on {Systems}, {Man}, and {Cybernetics} ({SMC})}}}\ (\bibinfo  {publisher} {IEEE},\ \bibinfo {address} {Honolulu, Oahu, HI, USA},\ \bibinfo {year} {2023})\ pp.\ \bibinfo {pages} {1058--1063}\BibitemShut {NoStop}%
\bibitem [{\citenamefont {Evans}\ \emph {et~al.}(2024)\citenamefont {Evans}, \citenamefont {Cook}, \citenamefont {Bradshaw},\ and\ \citenamefont {LaBorde}}]{evans_learning_2024}%
  \BibitemOpen
  \bibfield  {author} {\bibinfo {author} {\bibfnamefont {E.~N.}\ \bibnamefont {Evans}}, \bibinfo {author} {\bibfnamefont {M.}~\bibnamefont {Cook}}, \bibinfo {author} {\bibfnamefont {Z.~P.}\ \bibnamefont {Bradshaw}},\ and\ \bibinfo {author} {\bibfnamefont {M.~L.}\ \bibnamefont {LaBorde}},\ }\href {https://doi.org/10.48550/arXiv.2403.14753} {\bibinfo {title} {Learning with {SASQuaTCh}: a {Novel} {Variational} {Quantum} {Transformer} {Architecture} with {Kernel}-{Based} {Self}-{Attention}}} (\bibinfo {year} {2024}),\ \bibinfo {note} {arXiv:2403.14753 [quant-ph]}\BibitemShut {NoStop}%
\bibitem [{\citenamefont {Xue}\ \emph {et~al.}(2024)\citenamefont {Xue}, \citenamefont {Chen}, \citenamefont {Zhuang}, \citenamefont {Wang}, \citenamefont {Sun}, \citenamefont {Wang}, \citenamefont {Liu}, \citenamefont {Wu}, \citenamefont {Wang},\ and\ \citenamefont {Guo}}]{xue_end--end_2024}%
  \BibitemOpen
  \bibfield  {author} {\bibinfo {author} {\bibfnamefont {C.}~\bibnamefont {Xue}}, \bibinfo {author} {\bibfnamefont {Z.-Y.}\ \bibnamefont {Chen}}, \bibinfo {author} {\bibfnamefont {X.-N.}\ \bibnamefont {Zhuang}}, \bibinfo {author} {\bibfnamefont {Y.-J.}\ \bibnamefont {Wang}}, \bibinfo {author} {\bibfnamefont {T.-P.}\ \bibnamefont {Sun}}, \bibinfo {author} {\bibfnamefont {J.-C.}\ \bibnamefont {Wang}}, \bibinfo {author} {\bibfnamefont {H.-Y.}\ \bibnamefont {Liu}}, \bibinfo {author} {\bibfnamefont {Y.-C.}\ \bibnamefont {Wu}}, \bibinfo {author} {\bibfnamefont {Z.-L.}\ \bibnamefont {Wang}},\ and\ \bibinfo {author} {\bibfnamefont {G.-P.}\ \bibnamefont {Guo}},\ }\href {https://doi.org/10.48550/arXiv.2402.18940} {\bibinfo {title} {End-to-{End} {Quantum} {Vision} {Transformer}: {Towards} {Practical} {Quantum} {Speedup} in {Large}-{Scale} {Models}}} (\bibinfo {year} {2024}),\ \bibinfo {note} {arXiv:2402.18940 [quant-ph]}\BibitemShut {NoStop}%
\bibitem [{\citenamefont {Mitarai}\ \emph {et~al.}(2019)\citenamefont {Mitarai}, \citenamefont {Kitagawa},\ and\ \citenamefont {Fujii}}]{mitarai_quantum_2019}%
  \BibitemOpen
  \bibfield  {author} {\bibinfo {author} {\bibfnamefont {K.}~\bibnamefont {Mitarai}}, \bibinfo {author} {\bibfnamefont {M.}~\bibnamefont {Kitagawa}},\ and\ \bibinfo {author} {\bibfnamefont {K.}~\bibnamefont {Fujii}},\ }\href {https://doi.org/10.1103/PhysRevA.99.012301} {\bibfield  {journal} {\bibinfo  {journal} {Physical Review A}\ }\textbf {\bibinfo {volume} {99}},\ \bibinfo {pages} {012301} (\bibinfo {year} {2019})},\ \bibinfo {note} {arXiv:1805.11250 [quant-ph]}\BibitemShut {NoStop}%
\bibitem [{\citenamefont {Cherrat}\ \emph {et~al.}(2024)\citenamefont {Cherrat}, \citenamefont {Kerenidis}, \citenamefont {Mathur}, \citenamefont {Landman}, \citenamefont {Strahm},\ and\ \citenamefont {Li}}]{cherrat_quantum_2024}%
  \BibitemOpen
  \bibfield  {author} {\bibinfo {author} {\bibfnamefont {E.~A.}\ \bibnamefont {Cherrat}}, \bibinfo {author} {\bibfnamefont {I.}~\bibnamefont {Kerenidis}}, \bibinfo {author} {\bibfnamefont {N.}~\bibnamefont {Mathur}}, \bibinfo {author} {\bibfnamefont {J.}~\bibnamefont {Landman}}, \bibinfo {author} {\bibfnamefont {M.}~\bibnamefont {Strahm}},\ and\ \bibinfo {author} {\bibfnamefont {Y.~Y.}\ \bibnamefont {Li}},\ }\href {https://doi.org/10.22331/q-2024-02-22-1265} {\bibfield  {journal} {\bibinfo  {journal} {Quantum}\ }\textbf {\bibinfo {volume} {8}},\ \bibinfo {pages} {1265} (\bibinfo {year} {2024})}\BibitemShut {NoStop}%
\bibitem [{\citenamefont {Liao}\ and\ \citenamefont {Ferrie}(2024)}]{liao_gpt_2024}%
  \BibitemOpen
  \bibfield  {author} {\bibinfo {author} {\bibfnamefont {Y.}~\bibnamefont {Liao}}\ and\ \bibinfo {author} {\bibfnamefont {C.}~\bibnamefont {Ferrie}},\ }\href {https://doi.org/10.48550/arXiv.2403.09418} {\bibinfo {title} {{GPT} on a {Quantum} {Computer}}} (\bibinfo {year} {2024}),\ \bibinfo {note} {arXiv:2403.09418 [quant-ph]}\BibitemShut {NoStop}%
\bibitem [{\citenamefont {Guo}\ \emph {et~al.}(2024)\citenamefont {Guo}, \citenamefont {Yu}, \citenamefont {Choi}, \citenamefont {Agrawal}, \citenamefont {Nakaji}, \citenamefont {Aspuru-Guzik},\ and\ \citenamefont {Rebentrost}}]{guo_quantum_2024}%
  \BibitemOpen
  \bibfield  {author} {\bibinfo {author} {\bibfnamefont {N.}~\bibnamefont {Guo}}, \bibinfo {author} {\bibfnamefont {Z.}~\bibnamefont {Yu}}, \bibinfo {author} {\bibfnamefont {M.}~\bibnamefont {Choi}}, \bibinfo {author} {\bibfnamefont {A.}~\bibnamefont {Agrawal}}, \bibinfo {author} {\bibfnamefont {K.}~\bibnamefont {Nakaji}}, \bibinfo {author} {\bibfnamefont {A.}~\bibnamefont {Aspuru-Guzik}},\ and\ \bibinfo {author} {\bibfnamefont {P.}~\bibnamefont {Rebentrost}},\ }\href {https://doi.org/10.48550/arXiv.2402.16714} {\bibinfo {title} {Quantum linear algebra is all you need for {Transformer} architectures}} (\bibinfo {year} {2024}),\ \bibinfo {note} {arXiv:2402.16714 [quant-ph]}\BibitemShut {NoStop}%
\bibitem [{\citenamefont {Devlin}\ \emph {et~al.}(2019)\citenamefont {Devlin}, \citenamefont {Chang}, \citenamefont {Lee},\ and\ \citenamefont {Toutanova}}]{devlin_bert_2019}%
  \BibitemOpen
  \bibfield  {author} {\bibinfo {author} {\bibfnamefont {J.}~\bibnamefont {Devlin}}, \bibinfo {author} {\bibfnamefont {M.-W.}\ \bibnamefont {Chang}}, \bibinfo {author} {\bibfnamefont {K.}~\bibnamefont {Lee}},\ and\ \bibinfo {author} {\bibfnamefont {K.}~\bibnamefont {Toutanova}},\ }\href {https://doi.org/10.48550/arXiv.1810.04805} {\bibinfo {title} {{BERT}: {Pre}-training of {Deep} {Bidirectional} {Transformers} for {Language} {Understanding}}} (\bibinfo {year} {2019}),\ \bibinfo {note} {arXiv:1810.04805}\BibitemShut {NoStop}%
\bibitem [{\citenamefont {Abbas}\ \emph {et~al.}(2023)\citenamefont {Abbas}, \citenamefont {King}, \citenamefont {Huang}, \citenamefont {Huggins}, \citenamefont {Movassagh}, \citenamefont {Gilboa},\ and\ \citenamefont {McClean}}]{abbas_quantum_2023}%
  \BibitemOpen
  \bibfield  {author} {\bibinfo {author} {\bibfnamefont {A.}~\bibnamefont {Abbas}}, \bibinfo {author} {\bibfnamefont {R.}~\bibnamefont {King}}, \bibinfo {author} {\bibfnamefont {H.-Y.}\ \bibnamefont {Huang}}, \bibinfo {author} {\bibfnamefont {W.~J.}\ \bibnamefont {Huggins}}, \bibinfo {author} {\bibfnamefont {R.}~\bibnamefont {Movassagh}}, \bibinfo {author} {\bibfnamefont {D.}~\bibnamefont {Gilboa}},\ and\ \bibinfo {author} {\bibfnamefont {J.~R.}\ \bibnamefont {McClean}},\ }\href {https://doi.org/10.48550/arXiv.2305.13362} {\bibinfo {title} {On quantum backpropagation, information reuse, and cheating measurement collapse}} (\bibinfo {year} {2023}),\ \bibinfo {note} {arXiv:2305.13362 [quant-ph]}\BibitemShut {NoStop}%
\bibitem [{\citenamefont {Spall}(1992)}]{spall_multivariate_1992}%
  \BibitemOpen
  \bibfield  {author} {\bibinfo {author} {\bibfnamefont {J.}~\bibnamefont {Spall}},\ }\href {https://doi.org/10.1109/9.119632} {\bibfield  {journal} {\bibinfo  {journal} {IEEE Transactions on Automatic Control}\ }\textbf {\bibinfo {volume} {37}},\ \bibinfo {pages} {332} (\bibinfo {year} {1992})}\BibitemShut {NoStop}%
\bibitem [{\citenamefont {Hoffmann}\ and\ \citenamefont {Brown}(2022)}]{hoffmann_gradient_2022}%
  \BibitemOpen
  \bibfield  {author} {\bibinfo {author} {\bibfnamefont {T.}~\bibnamefont {Hoffmann}}\ and\ \bibinfo {author} {\bibfnamefont {D.}~\bibnamefont {Brown}},\ }\href {https://doi.org/10.48550/arXiv.2211.13981} {\bibinfo {title} {Gradient {Estimation} with {Constant} {Scaling} for {Hybrid} {Quantum} {Machine} {Learning}}} (\bibinfo {year} {2022}),\ \bibinfo {note} {arXiv:2211.13981 [quant-ph]}\BibitemShut {NoStop}%
\bibitem [{\citenamefont {Ramakrishnan}\ \emph {et~al.}(2014)\citenamefont {Ramakrishnan}, \citenamefont {Dral}, \citenamefont {Rupp},\ and\ \citenamefont {von Lilienfeld}}]{ramakrishnan_quantum_2014}%
  \BibitemOpen
  \bibfield  {author} {\bibinfo {author} {\bibfnamefont {R.}~\bibnamefont {Ramakrishnan}}, \bibinfo {author} {\bibfnamefont {P.~O.}\ \bibnamefont {Dral}}, \bibinfo {author} {\bibfnamefont {M.}~\bibnamefont {Rupp}},\ and\ \bibinfo {author} {\bibfnamefont {O.~A.}\ \bibnamefont {von Lilienfeld}},\ }\href {https://doi.org/10.1038/sdata.2014.22} {\bibfield  {journal} {\bibinfo  {journal} {Scientific Data}\ }\textbf {\bibinfo {volume} {1}},\ \bibinfo {pages} {140022} (\bibinfo {year} {2014})}\BibitemShut {NoStop}%
\bibitem [{\citenamefont {Weininger}(1988)}]{weininger_smiles_1988}%
  \BibitemOpen
  \bibfield  {author} {\bibinfo {author} {\bibfnamefont {D.}~\bibnamefont {Weininger}},\ }\href {https://doi.org/10.1021/ci00057a005} {\bibfield  {journal} {\bibinfo  {journal} {Journal of Chemical Information and Computer Sciences}\ }\textbf {\bibinfo {volume} {28}},\ \bibinfo {pages} {31} (\bibinfo {year} {1988})}\BibitemShut {NoStop}%
\bibitem [{\citenamefont {{Greg Landrum}}\ \emph {et~al.}(2024)\citenamefont {{Greg Landrum}}, \citenamefont {{Paolo Tosco}}, \citenamefont {{Brian Kelley}}, \citenamefont {{Ricardo Rodriguez}}, \citenamefont {{David Cosgrove}}, \citenamefont {{Riccardo Vianello}}, \citenamefont {{sriniker}}, \citenamefont {{Peter Gedeck}}, \citenamefont {{Gareth Jones}}, \citenamefont {{NadineSchneider}}, \citenamefont {{Eisuke Kawashima}}, \citenamefont {{Dan Nealschneider}}, \citenamefont {{Andrew Dalke}}, \citenamefont {{Matt Swain}}, \citenamefont {{Brian Cole}},\ and\ \citenamefont {{et al.}}}]{greg_landrum_rdkitrdkit_2024}%
  \BibitemOpen
  \bibfield  {author} {\bibinfo {author} {\bibnamefont {{Greg Landrum}}}, \bibinfo {author} {\bibnamefont {{Paolo Tosco}}}, \bibinfo {author} {\bibnamefont {{Brian Kelley}}}, \bibinfo {author} {\bibnamefont {{Ricardo Rodriguez}}}, \bibinfo {author} {\bibnamefont {{David Cosgrove}}}, \bibinfo {author} {\bibnamefont {{Riccardo Vianello}}}, \bibinfo {author} {\bibnamefont {{sriniker}}}, \bibinfo {author} {\bibnamefont {{Peter Gedeck}}}, \bibinfo {author} {\bibnamefont {{Gareth Jones}}}, \bibinfo {author} {\bibnamefont {{NadineSchneider}}}, \bibinfo {author} {\bibnamefont {{Eisuke Kawashima}}}, \bibinfo {author} {\bibnamefont {{Dan Nealschneider}}}, \bibinfo {author} {\bibnamefont {{Andrew Dalke}}}, \bibinfo {author} {\bibnamefont {{Matt Swain}}}, \bibinfo {author} {\bibnamefont {{Brian Cole}}},\ and\ \bibinfo {author} {\bibnamefont {{et al.}}},\ }\href {https://doi.org/10.5281/ZENODO.591637} {\bibinfo {title} {rdkit/rdkit: 2024\_09\_4 ({Q3} 2024) {Release}}} (\bibinfo {year} {2024})\BibitemShut {NoStop}%
\bibitem [{\citenamefont {Paszke}\ \emph {et~al.}(2019)\citenamefont {Paszke}, \citenamefont {Gross}, \citenamefont {Massa}, \citenamefont {Lerer}, \citenamefont {Bradbury}, \citenamefont {Chanan}, \citenamefont {Killeen}, \citenamefont {Lin}, \citenamefont {Gimelshein}, \citenamefont {Antiga}, \citenamefont {Desmaison}, \citenamefont {Köpf}, \citenamefont {Yang}, \citenamefont {DeVito}, \citenamefont {Raison},\ and\ \citenamefont {{et al.}}}]{paszke_pytorch_2019}%
  \BibitemOpen
  \bibfield  {author} {\bibinfo {author} {\bibfnamefont {A.}~\bibnamefont {Paszke}}, \bibinfo {author} {\bibfnamefont {S.}~\bibnamefont {Gross}}, \bibinfo {author} {\bibfnamefont {F.}~\bibnamefont {Massa}}, \bibinfo {author} {\bibfnamefont {A.}~\bibnamefont {Lerer}}, \bibinfo {author} {\bibfnamefont {J.}~\bibnamefont {Bradbury}}, \bibinfo {author} {\bibfnamefont {G.}~\bibnamefont {Chanan}}, \bibinfo {author} {\bibfnamefont {T.}~\bibnamefont {Killeen}}, \bibinfo {author} {\bibfnamefont {Z.}~\bibnamefont {Lin}}, \bibinfo {author} {\bibfnamefont {N.}~\bibnamefont {Gimelshein}}, \bibinfo {author} {\bibfnamefont {L.}~\bibnamefont {Antiga}}, \bibinfo {author} {\bibfnamefont {A.}~\bibnamefont {Desmaison}}, \bibinfo {author} {\bibfnamefont {A.}~\bibnamefont {Köpf}}, \bibinfo {author} {\bibfnamefont {E.}~\bibnamefont {Yang}}, \bibinfo {author} {\bibfnamefont {Z.}~\bibnamefont {DeVito}}, \bibinfo {author} {\bibfnamefont {M.}~\bibnamefont {Raison}},\ and\ \bibinfo {author} {\bibnamefont {{et al.}}},\ }\href
  {https://doi.org/10.48550/arXiv.1912.01703} {\bibinfo {title} {{PyTorch}: {An} {Imperative} {Style}, {High}-{Performance} {Deep} {Learning} {Library}}} (\bibinfo {year} {2019}),\ \bibinfo {note} {arXiv:1912.01703}\BibitemShut {NoStop}%
\bibitem [{\citenamefont {Loshchilov}\ and\ \citenamefont {Hutter}(2019)}]{loshchilov_decoupled_2019}%
  \BibitemOpen
  \bibfield  {author} {\bibinfo {author} {\bibfnamefont {I.}~\bibnamefont {Loshchilov}}\ and\ \bibinfo {author} {\bibfnamefont {F.}~\bibnamefont {Hutter}},\ }\href {https://doi.org/10.48550/arXiv.1711.05101} {\bibinfo {title} {Decoupled {Weight} {Decay} {Regularization}}} (\bibinfo {year} {2019}),\ \bibinfo {note} {arXiv:1711.05101}\BibitemShut {NoStop}%
\bibitem [{\citenamefont {{The CUDA-Q development team}}()}]{The_CUDA-Q_development_team_CUDA-Q}%
  \BibitemOpen
  \bibfield  {author} {\bibinfo {author} {\bibnamefont {{The CUDA-Q development team}}},\ }\href {https://github.com/NVIDIA/cuda-quantum} {\bibinfo {title} {{CUDA-Q}}}\BibitemShut {NoStop}%
\bibitem [{\citenamefont {Pedregosa}\ \emph {et~al.}(2011)\citenamefont {Pedregosa}, \citenamefont {Varoquaux}, \citenamefont {Gramfort}, \citenamefont {Michel}, \citenamefont {Thirion}, \citenamefont {Grisel}, \citenamefont {Blondel}, \citenamefont {Prettenhofer}, \citenamefont {Weiss}, \citenamefont {Dubourg},\ and\ \citenamefont {{others}}}]{pedregosa_scikit-learn_2011}%
  \BibitemOpen
  \bibfield  {author} {\bibinfo {author} {\bibfnamefont {F.}~\bibnamefont {Pedregosa}}, \bibinfo {author} {\bibfnamefont {G.}~\bibnamefont {Varoquaux}}, \bibinfo {author} {\bibfnamefont {A.}~\bibnamefont {Gramfort}}, \bibinfo {author} {\bibfnamefont {V.}~\bibnamefont {Michel}}, \bibinfo {author} {\bibfnamefont {B.}~\bibnamefont {Thirion}}, \bibinfo {author} {\bibfnamefont {O.}~\bibnamefont {Grisel}}, \bibinfo {author} {\bibfnamefont {M.}~\bibnamefont {Blondel}}, \bibinfo {author} {\bibfnamefont {P.}~\bibnamefont {Prettenhofer}}, \bibinfo {author} {\bibfnamefont {R.}~\bibnamefont {Weiss}}, \bibinfo {author} {\bibfnamefont {V.}~\bibnamefont {Dubourg}},\ and\ \bibinfo {author} {\bibnamefont {{others}}},\ }\href@noop {} {\bibfield  {journal} {\bibinfo  {journal} {Journal of machine learning research}\ }\textbf {\bibinfo {volume} {12}},\ \bibinfo {pages} {2825} (\bibinfo {year} {2011})}\BibitemShut {NoStop}%
\end{thebibliography}%


\appendix
\renewcommand{\thetable}{A\arabic{table}}
\setcounter{table}{0}

\section{Property Choices for Inference}

For all models (Quantum, Classical--eq, Classical), the ranking of property value selection methods that maximize valid and unique molecules ($V \times U$) is: Mean $>$ Mode $>$ Median, as shown in Table \ref{tab:appendix_validity}. Interestingly, this statistical choice affects novelty, with mode-based inference increasing the fraction of novel compounds to over 95\%. We tested the models’ ability to generate molecules beyond the training distribution using $\text{median} \pm 1.5$ IQR targets, as presented in Table \ref{tab:appendix_conditions}, to assess whether distributional skew disproportionately impacts any model. For the upper range, the Quantum model performed best for nRing; the Classical--eq model for MW, nHet, LogP, and Stereo; and the Classical model for HBA, HBD, nRot, and TPSA. For the lower range, the Quantum model performed best for HBA, nHet, and LogP; the Classical--eq model for TPSA; and the Classical model for MW, HBD, nRot, and Stereo. These results--given the margins observed--suggest no model outperforms or underperforms in generating target molecules when using $\mu \pm 2 \sigma$ versus $\text{median} \pm 1.5$ IQR approaches.

\begin{table*}[]
\caption{\label{tab:appendix_validity}%
Inference performance of each model at the epoch with the lowest validation loss, where the conditions used for generation are chosen from the Mean, Median, and Mode of those properties from within the training data. Quantum indicates the quantum-classical hybrid model, Classical -- eq denotes the fully classical model with an equal number parameters as the Quantum model, Classical denotes the fully classical model with an equivalent architecture to the Quantum model, but with traditionally sized weight matrices. Validity \% is the percentage of generated sequences that create a valid \texttt{mol} structure in RDKit out of 100,000 queries to the trained model. The product of validity (V) and uniqueness (U) shows the percentage of model queries which result in unique compounds. Novelty \% is the percentage of valid and unique SMILES strings that do not appear in the training set. 
}
\begin{ruledtabular}
\begin{tabular}{cccccccc}

&
\textrm{Model}
 &
\textrm{Validity \%}&
\textrm{Uniqueness \%}&
\textrm{V$\times$U \%}&
\textrm{Novelty \%}\\

\colrule
  & Quantum &  50.5 & 38.8 & 19.6 & 69.6\\
Mean & Classical -- eq &  50.7 & 40.0 & 20.3 & 70.4 \\
  & Classical  & 38.5 & 48.0 & 18.5 & 71.2\\
\colrule
  & Quantum  & 70.2 & 14.5 & 10.2 & 66.3\\
Median & Classical -- eq  & 80.3 & 14.5 & 11.6 & 66.3 \\
  & Classical  & 70.9 & 23.1 & 16.4 & 75.2\\
\colrule
  & Quantum  & 73.8 & 21.7 & 16.0 & 95.1\\
Mode & Classical -- eq  & 65.8 & 22.6 & 14.9 & 94.6 \\
  & Classical  & 64.9 & 28.0 & 18.2 & 95.4\\
\end{tabular}
\end{ruledtabular}
\end{table*}

\begin{table*}

\caption{\label{tab:appendix_conditions}%
Conditional generation results. The top section demonstrates how well each model is able to generate molecules targeting the median values of each property from the training data. The middle and lower sections indicate a target that is above and below the median value for each property by 1.5 $\times$ interquartile range (IQR), respectively. For each model, the average value for that property of all valid generated molecules are shown. In the middle and lower sections, each numerical entry represents the result from an inference experiment where only that property was specified and the remaining 8 properties were imputed from the training data with k-nearest neighbors. Bold values indicate which model generated molecules closer to the target value. Quantum indicates the quantum-classical hybrid model, Classical -- eq denotes the fully classical model with an equal number parameters as the Quantum model, Classical denotes the fully classical model with an equivalent architecture to the Quantum model, but with traditionally sized weight matrices. All inferences were performed with the epoch that possessed the lowest validation loss for each model.
}
\begin{ruledtabular}

\begin{tabular}{cccccccccc}
 &
\textrm{MW}&
\textrm{HBA}&
\textrm{HBD}&
\textrm{nRot}&
\textrm{nRing}&
\textrm{nHet}&
\textrm{TPSA}&
\textrm{logP}&
\textrm{Stereo}\\

\colrule
Median & 125.13 & 2.00 & 1.00 & 1.00 & 1.00 & 2.00 & 35.82 & 0.28 & 2.00 \\

Quantum & 122.51 & 1.93 & \textbf{1.24} & 0.70 & 1.00 & 1.94 & \textbf{35.58} & \textbf{0.24} & 2.17 \\
Classical -- eq & \textbf{124.51} & 1.95 & 1.26 & 0.81 & 1.00 & \textbf{1.97} & 34.89 & 0.33 & \textbf{2.05} \\
Classical & 123.55 & \textbf{1.96} & 0.70 & \textbf{0.84} & 1.00 & 2.04 & 31.51 & 0.53 & 1.58 \\

\colrule
Median + (1.5 $\times$ IQR) &  134.07 & 3.50 & 2.50 & 2.50 & 2.50 & 3.50 & 81.29 & 2.23 & 6.50 \\

Quantum & 131.98 & 3.08 & 2.61 & 2.05 & \textbf{2.39} & 3.23 & 71.37 & 2.14 & 5.29 \\
Classical -- eq & \textbf{134.31} & 3.32 & 2.96 & 1.99 & 2.25 & \textbf{3.37} & 86.42 & \textbf{2.27} & \textbf{5.32} \\
Classical & 134.66 & \textbf{3.50} & \textbf{2.50} & \textbf{2.29} & 2.12 & 3.70 & \textbf{81.63} & 2.47 & 5.12 \\

\colrule
Median - (1.5 $\times$ IQR) & 116.19 & 0.50 & 0.00 & 0.00 & 0.00 & 0.50 & 0.00 & -1.66 & 0.00 \\
 
Quantum & 116.09 & \textbf{1.07} & 0.23 & 0.15 & 0.00 & \textbf{0.45} & 0.92 & \textbf{-1.60} & 0.21 \\
Classical -- eq & 119.34 & 1.12 & 0.48 & 0.13 & 0.00 & 0.77 & \textbf{0.87} & -1.58 & 0.24 \\
Classical & \textbf{116.22} & 1.14 & \textbf{0.16} & \textbf{0.11} & 0.00 & 0.81 & 1.02 & -1.41 & \textbf{0.10} \\

\end{tabular}
\end{ruledtabular}
\end{table*}

\


\end{document}